\documentclass[amsmath,eqsecnum,preprint,pre,aps,showpacs]{revtex4}
\usepackage{amsmath}
\usepackage{graphicx}

\begin{document}

\title{General Non-equilibrium Theory of Colloid Dynamics }
\author{ Pedro Ram\'irez-Gonz\'alez and Magdaleno Medina-Noyola}

\address{Instituto de F\'{\i}sica {\sl ``Manuel Sandoval
Vallarta"}, Universidad Aut\'{o}noma de San Luis Potos\'{\i},
\'{A}lvaro Obreg\'{o}n 64, 78000 San Luis Potos\'{\i}, SLP,
M\'{e}xico}
\date{\today}

\begin{abstract}

A non-equilibrium extension of Onsager's canonical theory of
thermal fluctuations is employed to derive a self-consistent
theory for the description of the statistical properties of
the instantaneous local concentration profile $n(\textbf{r},t)$
of a colloidal liquid in terms of the coupled time evolution
equations of its mean value $\overline{n}(\textbf{r},t)$ and
of the covariance $\sigma(\textbf{r},\textbf{r}';t)\equiv
\overline{\delta n(\textbf{r},t)\delta n (\textbf{r}',t)}$ of
its fluctuations $\delta n(\textbf{r},t) = n(\textbf{r},t)-
\overline{n}(\textbf{r},t)$. These two coarse-grained equations
involve a local mobility function $b(\textbf{r},t)$ which, in
its turn, is written in terms of the memory function of the
two-time correlation function $C(\textbf{r},\textbf{r}';t,t')
\equiv \overline{\delta n(\textbf{r},t)\delta n (\textbf{r}',t')}$.
For given effective interactions between colloidal particles and
applied external fields, the resulting self-consistent theory is
aimed at describing the evolution of a strongly correlated colloidal
liquid from an initial state with arbitrary mean and covariance
$\overline{n}^0(\textbf{r})$ and $\sigma^0(\textbf{r},\textbf{r}')$
towards its equilibrium state characterized by the equilibrium local
concentration profile $\overline{n}^{eq}(\textbf{r})$ and
equilibrium covariance $\sigma^{eq}(\textbf{r},\textbf{r}')$.

This theory also provides a general theoretical framework to
describe irreversible processes associated with dynamic arrest
transitions, such as aging, and the effects of spatial
heterogeneities.

\bigskip
\bigskip
\bigskip

\end{abstract}


\maketitle

\section{Introduction}\label{I}

In this paper a non-equilibrium generalization is presented of the
self-consistent generalized Langevin equation (SCGLE) theory of
colloid dynamics \cite{scgle1,marco2}, and of its recent adaptation
as a theory of dynamic arrest \cite{rmf,todos2}, with the purpose of
describing non-equilibrium diffusive phenomena in general, and
irreversible aging processes associated with the glass and the gel
transitions \cite{pham1,cipelletti1,martinezvanmegen,sanz1,lu1} in
particular. This generalized theory is based on a non-equilibrium
extension of Onsager's canonical theory of thermal fluctuations. The
resulting theory contains, for example, the fundamental equation of
dynamic density functional theory \cite{tarazona1} as a particular
limit, whereas in other limit one can recognize the basic equation
of the theory of early spinodal decomposition \cite{langer}. A
practical and concrete use of the resulting general theory of
colloid dynamics is illustrated in a related paper \cite{aging2}
with a quantitative application to the prediction of the aging
processes occurring in a suddenly quenched colloidal liquid.

The dynamic properties of colloidal dispersions has been the subject
of sustained interest for many years \cite{1,6,2}. These properties
can be described in terms of the relaxation of the fluctuations
$\delta n({\bf r} ,t)$ of the local concentration $n({\bf r},t)$ of
colloidal particles around its bulk equilibrium value $n=N/V$. The
average decay of $\delta n({\bf r},t)$ is described by the two-time
correlation function $F(k,\tau;t)\equiv V^{-1} \overline{ \delta
n({\bf k},t+\tau)\delta n(-{\bf k},t)}$ of the Fourier transform
$\delta n({\bf k},t)$ of the fluctuations $\delta n({\bf r} ,t)$,
whose equal-time limit is $S(k;t)\equiv F(k,\tau=0;t) =V^{-1}
\overline{ \delta n({\bf k},t)\delta n(-{\bf k},t)} $. We shall
refer to the time $\tau$ as the \emph{correlation time}. If some
external (or internal) constraints that kept a system at a certain
macroscopic state are broken at the (\emph{evolution}) time $t=0$
the system relaxes spontaneously, searching its new thermodynamic
equilibrium state. If the end state, however, is a glass or a gel,
one refers to $t$ as the \emph{waiting} or \emph{aging} time
\cite{pham1,cipelletti1,martinezvanmegen,sanz1,lu1}. The evolution
of $S(k;t)$ and $F(k,\tau;t)$ as a function of the time $t$
characterizes the non-equilibrium evolution of the system, and its
theoretical understanding is a major fundamental challenge.

If the system is a fluid and it has fully relaxed to its
thermodynamic equilibrium state, the properties above no longer
depend on $t$, i.e., $F(k,\tau;t)=F(k,\tau)$ and $S(k;t)=S(k)$. The
equilibrium stationary correlation function $F(k,\tau)$ is then
referred to as the intermediate scattering function, and its initial
value $S(k)$ as the \emph{equilibrium} static structure factor.
These properties can be measured by a variety of experimental
techniques, including (static and/or dynamic) light scattering
\cite{cipelletti1,martinezvanmegen,1}. $S(k)$, being an equilibrium
property, is amenable to theoretical calculation using statistical
thermodynamic methods \cite{mcquarrie}. The fundamental
understanding of $F(k,\tau)$, on the other hand, requires the
development of theoretical methods to describe the correlations of
the local concentration fluctuations, and a number of such
approaches have been proposed for their theoretical calculation
\cite{1,6,2,13}. One of them has been developed within the last
decade and is referred to as the self-consistent generalized
Langevin equation (SCGLE) theory of colloid dynamics
\cite{scgle0,scgle1,scgle2,marco1,marco2}. This theory has been
recently applied to the description of dynamic arrest phenomena in
several specific colloidal systems that include mono-disperse
suspensions with hard-sphere interactions, moderately soft-sphere
and electrostatic repulsions, short-ranged attractive interactions,
and model mixtures of neutral and charged particles
\cite{rmf,todos1,todos2,attractive1,soft1,rigo1,rigo2,luis1}.

In spite of the long tradition in the study of glasses
\cite{angell,debenedetti,goetze1}, until recently the only
well-established and successful theoretical framework leading to
\emph{first-principles quantitative} predictions of the dynamic
properties of colloidal liquids near their dynamic arrest transition
was the conventional mode coupling theory (MCT) of the ideal glass
transition \cite{goetze1,goetze2,goetze3,goetze4}. Many of the
predictions of this theory have been systematically confirmed by
their detailed comparison with experimental measurements in model
colloidal systems \cite{vanmegen1, bartsch, beck, chen1, chen2,pham,
sciortinotartaglia,buzzacaro}. In this context, we can mention that
the more recently-developed SCGLE theory of dynamic arrest leads to
similar dynamic arrest scenarios as MCT \cite{rmf, todos1} for
several specific (mostly mono-disperse) systems, although for
colloidal mixtures differences may appear in some circumstances, as
reported in Refs. \cite{rigo1,rigo2,luis1}.

An important common feature of both theories in their current status
is that they are able to predict the regions of the control
parameter space in which the system is expected to be dynamically
arrested, i.e., they predict what we refer to as the ``dynamic
arrest phase diagram" of the system \cite{rigo2,luis1}. While it is
important to pursue the application of these two theories to
specific idealized or experimental model systems and to compare
their predictions, it is also important to attempt their extension
to the description of the detailed non-equilibrium processes leading
to dynamically arrested end states. Aging effects, for example,
should be a fundamental aspect of the experimental and theoretical
characterization of these non-equilibrium states. These
preoccupations have been addressed in the field of spin glasses,
where a mean-field theory has been developed within the last two
decades \cite{cugliandolo1}. The models involved, however, lack a
geometric structure and hence cannot describe the spatial evolution
of real colloidal glass formers. Although experimental studies
\cite{pham1,cipelletti1,martinezvanmegen,sanz1,lu1} and computer
simulations \cite{kobbarrat,foffiaging1,puertas1} have provided
important information about general properties of aging, until now
no quantitative first-principles theory is available to describe the
irreversible formation of structural glasses.

About a decade ago Latz \cite{latz} attempted to extend MCT to
describe the irreversible relaxation, including  aging processes, of
a suddenly quenched glass forming system. A major aspect of his work
involved the generalization to non-equilibrium conditions of the
conventional equilibrium projection operator approach \cite{berne}
to derive the corresponding memory function equations in which the
mode coupling approximations could be introduced. Similarly, De
Gregorio et al. \cite{degregorio} discussed time-translational
invariance and the fluctuation-dissipation theorem in the context of
the description of slow dynamics in system out of equilibrium but
close to dynamical arrest. They also proposed extensions of
approximations long known within MCT. Unfortunately, in neither of
these two theoretical efforts, quantitative predictions were
presented that could be contrasted with experimental or simulated
results in specific model systems of structural glass-formers.

The present work is aimed at extending the SCGLE theory of dynamic
arrest to non-equilibrium conditions. This paper contains the
proposal of such general theory while the accompanying paper (paper
II) reports a concrete quantitative application. The general theory
proposed here consists of the time evolution equations for the mean
value and for the covariance of the instantaneous local
concentration profile $n(\textbf{r},t)$ of a colloidal liquid
coupled, through a local mobility function $b(\textbf{r},t)$, with
two-time correlation function $C(\textbf{r},\textbf{r}';t,t')\equiv
\overline{\delta n (\textbf{r},t)\delta n (\textbf{r}',t')}$ of the
local concentration fluctuations. A set of well-defined
approximations in the memory function of
$C(\textbf{r},\textbf{r}';t,t')$ leads to the non-equilibrium
extension of the self-consistent generalized Langevin equation
theory of colloid dynamics to spatially non-uniform and temporally
non-stationary systems. The resulting theory is applied in II to the
description of aging effects in a specific model glass forming
colloidal liquid.

In contrast with MCT, the SCGLE theory does not involve the
assumption of an underlying Hamiltonian (or any other microscopic)
level of description, nor the use of projection operator techniques.
Instead, it is based on what we refer to as Onsager's canonical
theory of equilibrium thermal fluctuations. Since the description of
thermal fluctuations and relaxation processes can be approached from
a bewildering number of theoretical perspectives, involving a
diversity of issues, approaches, aims, methodologies, and
nomenclature \cite{keizer,degrootmazur,casasvazquez0}, it is
necessary to state that in this work for ``Onsager's theory" we mean
the general and fundamental laws of linear irreversible
thermodynamics and the corresponding stochastic theory of thermal
fluctuations, as stated by Onsager \cite{onsager1, onsager2} and by
Onsager and Machlup \cite{onsagermachlup1,onsagermachlup2},
respectively, with an adequate extension \cite{delrio,faraday} to
allow for the description of memory effects.

Viewed as a theory of fluctuations, Onsager's theory refers to
systems in thermodynamic equilibrium, and hence, assumes stationary
conditions. Thus, generalizing the SCGLE theory of colloid dynamics
to non-equilibrium calls for an extension of Onsager's theory to
non-stationary non-equilibrium conditions, outside the so called
``linear regime", where its validity has been universally tested
\cite{keizer}. Such an extended Onsager's theory is discussed in
detail elsewhere \cite{generalizedonsager}, and here we only provide
a brief summary (see Sec. \ref{II} below). In essence, however, this
extension consists of the assumption that the $t$-dependent
irreversible evolution of a system towards its stable equilibrium
state proceeds as a virtually continuous sequence of
non-equilibrium, but momentarily stationary, states. The main
objective of the present paper is then to apply this extended
canonical theory as a fundamental framework in which to discuss the
dynamics of a colloidal suspension that evolves irreversibly towards
its equilibrium state. Such application is the subject of Sec.
\ref{III}.

According to this program, our discussion will involve two distinct
levels of generality. The first corresponds to the rather abstract
and most general description provided by Onsager's extended theory
in terms of a set of macroscopic state variables, generically
denoted by $(a_1,a_2,...,a_M)\equiv \textbf{a}$, as reviewed in the
following section. The second corresponds to the description of
diffusive processes in colloidal dispersions, where the abstract
objects in Onsager's theory take a concrete meaning. Bridging these
two levels of discussion requires that we identify the specific
correspondence between the abstract concepts in Onsager's theory and
the concrete concepts pertaining to the other more specific level.
For example, the abstract state variables $a_i$ will be identified
with $N_i/ \Delta V$, the number concentration of particles in the
$i$th cell of an imaginary partitioning of the volume occupied by
the colloidal system in $M$ cells of volume $\Delta V$. In the
continuum limit, the components of the state vector $\textbf{a}(t)$
then become the local concentration profile $n(\textbf{r},t)$ and
the \emph{fundamental thermodynamic relation} $S=S[\textbf{a}]$
(which assigns to any point \textbf{a} of the thermodynamic state
space a value of the entropy \cite{callen}) will be identified with
the functional dependence of the free energy on the local
concentration profile employed, for example, in the classical
density functional theory \cite{evans} or in its more recent dynamic
version \cite{tarazona1,archer,royalvanblaaderen}. For completeness,
the structure of this thermodynamic framework is reviewed in an
Appendix. Finally, in the last section we summarize the main
conclusions of the present work.

\section{Generalized Onsager theory}\label{II}

In this section we summarize the main features of the extension of
Onsager's theory to non-stationary non-equilibrium states presented
in detail in Ref. \cite{generalizedonsager}. Thus, consider a system
whose macroscopic state is described in terms of a set of $M$
extensive variables $ a_i(t)$, $i=1,2,...,M$, which we group as the
components of a $M$-component (column) vector $\textbf{a}(t)$. The
fundamental postulate of this generalized theory is that the
dynamics of the state vector $\textbf{a}(t)$ may be represented by a
\emph{multivariate stochastic process} which is globally
non-stationary, but that within any small interval of the evolution
time $t$ may be regarded as approximately stationary. This local
stationarity approximation is then complemented with the assumption
that the mean value ${\overline {\textbf a }}(t)$ is the solution of
a generally nonlinear equation, represented by
\begin{equation}
\frac{d{\overline {\textbf a }}(t)}{dt}= \mathcal{R}\left[{\overline
{\textbf a }}(t)   \right],  \label{releq0}
\end{equation}
whose linear version in the deviations $\Delta\overline {{\textbf a
}}(t) \equiv \overline {{\textbf a }}(t) -{\textbf a }^{eq}$ from an
equilibrium  value ${\textbf a }^{eq}$ reads
\begin{equation}
\frac{d \Delta{\overline {\textbf a }}(t)}{dt}=
-\mathcal{L}[{\textbf a }^{eq}] \cdot \mathcal{E}[{\textbf a }^{eq}]
\cdot \Delta{\overline {\textbf a }}(t) , \label{releq0linearized}
\end{equation}
with $\mathcal{L}$ and $\mathcal{E}$ being $M$x$M$ matrices and with
the symbol ``$\cdot$" indicating the corresponding matrix product.
The  matrix $\mathcal{L} [{\textbf a}^{eq}]$ is referred to as the
kinetic matrix, related with the vector of ``fluxes"
$\mathcal{R}\left[{\textbf a }^{eq} \right]$ of Eq. (\ref{releq0})
by $\mathcal{L} [{\textbf a}^{eq}]\equiv -\left(
\partial \mathcal{R}\left[
{\textbf a }\right]/ \partial {\textbf a
}\right)_{\textbf{a}=\textbf{a}^{eq}} \cdot
\mathcal{E}^{-1}\left[{\textbf a}^{eq}\right]$.

On the other hand, $\mathcal{E}\left[{\textbf a }\right]$ is the
thermodynamic matrix, defined as
\begin{equation}
\mathcal{E}_{ij}[{\textbf a }] \equiv -\frac{1}{k_B}\left(
\frac{\partial^2 S[{\textbf a }]}{\partial a_i\partial a_j} \right)=
-\left( \frac{\partial F_i[{\textbf a }]}{\partial a_j} \right) \ \
\ \ (i,j=1,2,...,M), \label{matrixE}
\end{equation}
in which the function $S[{\textbf a }]$ determines the dependence of
the entropy on the components of the  vector $ {\textbf a}$, i.e.,
$S=S[{\textbf a }]$ is the fundamental thermodynamic relation of the
system, and hence,  $ F_j[ {\textbf a}] \equiv k_B^{-1} \left(
\partial S[ {\textbf a}] /
\partial a_j \right)$ is the conjugate intensive variable associated with $a_j$.
One should notice that Eq. (\ref{releq0linearized}) can be written
as $d\Delta {\textbf a }(t)/dt= \mathcal{L}[\textbf{a}^{eq}]\cdot
\Delta {\textbf F }(t)$, where $\Delta {\textbf F }(t)\equiv
{\textbf F [{\overline {\textbf a }}(t)]}-{\textbf F^{eq}  }$ is the
macroscopic deviation of the vector $ {\textbf F [{\overline
{\textbf a }}(t)]}$ of intensive parameters from its equilibrium
value ${\textbf F}^{eq} ={\textbf F [ {\textbf a }^{eq} ]}$. This
relaxation equation is immediately recognized as the classical
format of the linear laws of irreversible thermodynamics.

From these premises a time-evolution equation for the $M$x$M$
covariance matrix $\sigma (t) \equiv \overline{ \delta{\textbf a
}(t)\delta{\textbf a }^\dagger(t)}$ can be derived
\cite{generalizedonsager}, which reads
\begin{equation}
\frac{d\sigma(t)}{dt} = -\mathcal{L} [{\overline {\textbf a }}(t)]
\cdot \mathcal{E}\left[{\overline {\textbf a }}(t)\right]\cdot
\sigma (t) - \sigma (t) \cdot \mathcal{E}\left[{\overline{\textbf
a}}(t)\right] \cdot\mathcal{L}^{\dagger } [{\overline{\textbf
a}}(t)] + \left( \mathcal{L} [{\overline{\textbf a}}(t)] +
\mathcal{L}^{\dagger } [{\overline{\textbf a}}(t)]
\right).\label{sigmadtirrev2}
\end{equation}
This equation may be regarded as a simple extension of the equation
of motion for the covariance involved in the conventional Onsager
theory (see, for example, Eq. (1.8.9) of Ref. \cite{keizer}), in
which the matrices $\mathcal{L} [{\overline {\textbf a }}^{eq}]$ and
$ \mathcal{E}\left[{\overline {\textbf a }}^{eq}\right]$ are
replaced by $\mathcal{L} [{\overline {\textbf a }}(t)]$ and
$\mathcal{E}\left[{\overline {\textbf a }}(t)\right]$. The detailed
arguments to see that this is the proper manner to extend Onsager's
result to non-stationary processes can be found in Ref.
\cite{generalizedonsager}.

Thus, if two essential pieces of information were available, namely,
the fundamental thermodynamic relation $S=S[\textbf{a}]$ and the
state-dependence of $\mathcal{R}\left[ {\textbf a }\right]$, then
Eqs. (\ref{releq0}) and (\ref{sigmadtirrev2}) would constitute a
closed system of equations for the mean value $ {\overline {\textbf
a }}(t)$ and the covariance $\sigma (t)$. These are essentially the
first and second moments of the 1-time probability distribution
$P_1({\textbf a }_1,t_1)$ that the state vector {\textbf a } has the
value ${\textbf a }_1$ at the time $t=t_1$. Let us stress that the
full knowledge of this probability distribution is equivalent to the
full knowledge of the macroscopic state of the system, even though
for many purposes one may only be interested in one or some of its
moments. For example, the probability distribution of the
thermodynamic equilibrium state is fully determined by the
Boltzmann-Planck postulate to be \cite{keizer,landaulifshitz}
$P_1({\textbf a },t) =P^{eq}[{\bf a}]=\exp\left[{\left(S[{\bf
a}]-S[{\bf a}^{eq}]\right)/k_B}\right]$, whose mean value and
covariance are \cite{callen,greenecallen} $ {\overline {\textbf a
}}(t)={\textbf a }^{eq}$ and $\sigma (t)=\sigma^{eq} =
\mathcal{E}^{-1}\left[{\textbf a}^{eq}\right]$. One may stretch this
concept, and introduce the assumption that the non-equilibrium
evolution of the system can be described approximately by
$P_1({\textbf a },t) = P^{(l.e.)}({\textbf a },t) \equiv \exp
(S[{\textbf a }]-S[{\overline{\textbf a }}(t)])/k_B$, whose
covariance is given by $\sigma (t)=\sigma^{(l.e.)}  (t)=
\mathcal{E}^{-1}\left[{\overline {\textbf a }}(t)\right]$. We shall
refer to this as the \emph{local equilibrium approximation}, and an
idealized time-dependent process that satisfies this approximation
at any time $t$ shall be referred to as a \emph{quasi-static
process}. We must emphasize that the present extended Onsager's
theory is, of course, \emph{NOT} based on this approximation.

If the goal were to fully determine $P_1({\textbf a },t)$, in
principle, one could attempt to write the time-evolution equations
corresponding to the higher-order moments, thus constructing an
infinite hierarchy of equations for all such moments. Alternatively,
also in principle, one could attempt to write the time-evolution
equation for $P_1({\textbf a },t)$, from which one could determine
the time-evolution of all the moments. Our intention, however, is
not to follow any of these strategies, nor to assume that the
stochastic process is Gaussian, so that the first two moments above
will suffice to fully determine $P_1({\textbf a },t)$. In fact, we
are not actually interested in determining $P_1({\textbf a },t)$ at
all. Instead, our aim is to use the two general equations above for
$ {\overline {\textbf a }}(t)$ and $\sigma (t)$ as the fundamental
framework in which to introduce approximations that lead to a closed
system of equations for these directly measurable properties, at
least in specific and concrete cases, as in the colloidal context
discussed below. For this purpose, rather than analyzing the
higher-order moments of $P_1({\textbf a },t)$, we consider the
two-time correlation function, i.e., the second moment of the
two-time probability distribution $P_2({\textbf a }_1,t_1;{\textbf a
}_2,t_2)$, as one aspect of the properties of the thermal
fluctuations $\delta {\textbf a }(t+\tau)= {\textbf a
}(t+\tau)-\overline{{\textbf a }}(t)$ around the non-stationary mean
value $\overline{{\textbf a }}(t)$, within the local stationarity
approximation.

Thus, the second fundamental postulate of the generalized Onsager
theory is that the locally stationary fluctuations $\delta {\textbf
a }(t+\tau)$ can be described by a mathematical model that we refer
to as a \emph{generalized} Ornstein-Uhlenbeck process, discussed in
Ref. \cite{delrio}, which in the present context is defined by the
most general linear stochastic differential equation with additive
noise, which has the following general structure
\begin{equation}
\frac{d\delta  {\textbf a }(t +\tau)}{d\tau}= -
\omega[\overline{\textbf{a}}(t)] \cdot [ \sigma(t)]^{-1} \cdot
\delta {\textbf a }(t +\tau) - \int_0^{\tau}d\tau'
\gamma[\tau-\tau';\overline{\textbf{a}}(t)]\cdot [
\sigma(t)]^{-1}\cdot \delta {\textbf a }(t +\tau')+{\textbf f }(t
+\tau), \label{fluctuations1}
\end{equation}
in which the stochastic vector $\textbf{f}(t +\tau)$ is assumed
stationary but not necessarily Gaussian or $\delta$-correlated, the
matrix $\omega[\textbf{a}]$ is antisymmetric,
$\omega[\textbf{a}]=-\omega ^{ \dagger}[\textbf{a}]$, and the memory
matrix $\gamma[\tau;\textbf{a}]$ satisfies the
fluctuation-dissipation relation $ \gamma[\tau;\textbf{a}(t)] =
\gamma^{ \dagger}[-\tau;\textbf{a}(t)] =\  <{{\textbf f }(t
+\tau){\textbf f }^{\ \dagger}(t +0)}>$. From this generalized
Langevin equation one then derives the time-evolution equation for
the non-stationary time-correlation function $C(\tau;t)\equiv
\overline{\delta {\textbf a }(t+\tau)\delta {\textbf a }^{
\dagger}(t)}$, which reads
\begin{equation}
\frac{\partial C(\tau;t)}{\partial \tau}= -
\omega[\overline{\textbf{a}}(t)]\cdot \sigma^{-1} (t)\cdot C(\tau;t)
-\int_0^\tau d\tau ' \gamma[\tau-\tau';\overline{\textbf{a}}(t)]
\cdot \sigma^{-1}(t)\cdot C(\tau';t),  \label{fluctuations3}
\end{equation}
and whose initial condition is $C(\tau=0;t)=\sigma(t)$. This
equation describes the decay of the correlation function $C(\tau;t)$
with the ``microscopic" correlation time $\tau$, after the system
has evolved during a ``macroscopic" evolution time $t$ from an
initial state described by $ \textbf{a}^0 \equiv
\overline{\textbf{a}}(t=0)$ and $\sigma^0\equiv \sigma(t=0)$, to the
``current" state described by $\overline{\textbf{a}}(t)$ and
$\sigma(t)$.

Notice that this equation involves $\sigma(t)$ explicitly, and
$\overline{\textbf{a}}(t)$ implicitly through the matrices
$\omega[\overline{\textbf{a}}(t)]$ and
$\gamma[\tau;\overline{\textbf{a}}(t)]$. Thus, besides requiring the
actual values of $\overline{\textbf{a}}(t)$ and $\sigma(t)$, this
equation also requires the information on the state dependence of
the matrices $\omega[\textbf{a}]$ and $\gamma[\tau;\textbf{a}]$.
This information, however, must be closely related with the kinetic
matrix $\mathcal{L} [{\textbf a}]$. To establish such relationship,
notice that if the system has reached a thermodynamic equilibrium
state, in which $P_1({\textbf a },t) =P^{eq}[{\textbf a }]$, $
{\overline {\textbf a }}(t)={\textbf a }^{eq}$, and $\sigma
(t)=\sigma^{eq} = \mathcal{E}^{-1}\left[{\textbf a}^{eq}\right]$,
then Eq. (\ref{fluctuations1}) may be written, in the so-called
``Markov" limit, as

\begin{equation}
\frac{d\delta  {\textbf a }(\tau)}{d\tau}=
-\mathcal{L}[\textbf{a}^{eq}] \cdot \mathcal{E}\left[{\textbf
a}^{eq}\right]  \cdot \delta {\textbf a }(\tau) +{\textbf f }(\tau),
\label{fluctuationsOM}
\end{equation}
with $\mathcal{L}[\textbf{a}] $ defined as
\begin{equation}
\mathcal{L}[\textbf{a}] \equiv \omega[\textbf{a}] + \int_0^{\infty}
d\tau \gamma[\tau;\textbf{a}]. \label{okcssppp}
\end{equation}
Eq. (\ref{fluctuationsOM}), however, is the linear stochastic
equation with additive white noise of the Onsager-Machlup theory of
equilibrium fluctuations \cite{onsagermachlup1,onsagermachlup2}.
According to Onsager's regression hypothesis, this equation must be
identical, except for the additive white noise ${\textbf f }(\tau)$,
to the phenomenological relaxation equation in Eq.
(\ref{releq0linearized}). This requires that the definition of the
matrix $\mathcal{L}[\textbf{a}] $ in Eq. (\ref{okcssppp}) above must
be consistent with the phenomenological definition $\mathcal{L}
[{\textbf a}^{eq}]\equiv -\left(
\partial \mathcal{R}\left[
{\textbf a }\right]/ \partial {\textbf a
}\right)_{\textbf{a}=\textbf{a}^{eq}} \cdot
\mathcal{E}^{-1}\left[{\textbf a}^{eq}\right]$ beneath Eq.
(\ref{releq0linearized}).

The results above then state that the kinetic matrix
$\mathcal{L}[\textbf{a}]$ may be obtained either by linearizing the
non-linear phenomenological relaxation equation (\ref{releq0}) if
this equation is known a priori, or by means of the general
relationship in Eq. (\ref{okcssppp}) if the matrices
$\omega[{\overline {\textbf a }}(t)]$ and $\gamma[\tau;{\overline
{\textbf a }}(t)]$ can be determined by independent arguments, as we
propose here in the context of colloid dynamics. In general, the
antisymmetric matrix $\omega[{\overline {\textbf a }}(t)]$
represents conservative (mechanical, geometrical, or streaming)
terms, and its determination in specific contexts is relatively
straightforward. In contrast, the memory matrix
$\gamma[\tau;{\overline {\textbf a }}(t)]$ summarizes the effects of
all the complex dissipative irreversible processes taking place in
the system. Its exact determination is perhaps impossible except in
specific cases or limits; otherwise one must resort to
approximations. These may have the form of a closure relation
expressing $\gamma[\tau;{\overline {\textbf a }}(t)]$ in terms of
the two-time correlation matrix $C(\tau;t)$ itself, giving rise to a
self-consistent system of equations, as we illustrate in the
application that follows.

As a final observation, let us mention that throughout the previous
discussion we have assumed that the variables $a_i(t)$ represent
extensive state variables. In reality, one could also describe the
state of the system in terms of any combination of extensive and
intensive variables. The choice depends on the convenience, given
the macroscopic conditions imposed on the system. Using only
extensive variables, for example, is the most convenient choice if
the system is subject to isolation conditions. If, however, the
system is in contact with a thermal reservoir, the temperature,
rather than the internal energy, may be a more convenient variable.
On the other hand, if the external constraints (isolation, contact
with thermal reservoirs, applied external fields, etc.) are
time-independent, the time-evolution equations for the mean value
$\overline {\textbf a }(t)$, for the covariance $\sigma(t)$ and for
the correlation function $C(\tau;t)$ (Eqs. (\ref{releq0}),
(\ref{sigmadtirrev2}), and (\ref{fluctuations3})) will describe the
spontaneous relaxation of the system toward the corresponding
equilibrium state. We may, however, also consider the possibility
that these constraints vary in time in a programmed manner. In this
case, the time-evolution of the parameters describing these
constraints (for example, the overall density of the system or the
temperature of the heat reservoir) may be prescribed, rather than
determined as the solution of any time-evolution equation such as
Eq. (\ref{releq0}), and the time-evolution equations for the
covariance and the correlation function (Eqs. (\ref{sigmadtirrev2})
and (\ref{fluctuations3})) will describe the non-equilibrium
response of the system to these forced time-dependent macroscopic
constraints.

\section{Application to colloid dynamics}\label{III}

In this section we discuss the general problem of the diffusive
relaxation of the local concentration of colloidal particles in the
absence of hydrodynamic interactions but which interact through
pairwise direct forces represented by the effective pair potential
$u(\textbf{r}, \textbf{r}')$. Thus, let us consider a dispersion of
$N$ such colloidal particles of mass $m$ in a volume $V$ which, in
the absence of external fields, has a uniform bulk number
concentration $n_B = N/V$. In the presence of a conservative static
external field that exerts a force $\textbf{F}^{ext} (\textbf{r}) =
- \nabla \psi (\textbf{r}) $ on one particle located at position
$\textbf{r}$, the mean local concentration profile of colloidal
particles, $\overline{n} (\textbf{r},t)$, will evolve in time from
some initial condition $\overline{n }(\textbf{r},t=0)=n^0
(\textbf{r})$, towards its stable thermodynamic equilibrium value
$n^{eq} (\textbf{r})$, while the covariance
$\sigma(\textbf{r},\textbf{r}';t)\equiv \overline{\delta
n(\textbf{r},t)\delta n(\textbf{r}',t)}$ of the fluctuations $\delta
n(\textbf{r},t)\equiv n(\textbf{r},t)-\overline{n }(\textbf{r},t)$
will evolve from an initial value $\sigma^0(\textbf{r},\textbf{r}')$
to a final equilibrium value $\sigma^{eq}(\textbf{r},\textbf{r}')$.
The initial values $n^0 (\textbf{r})$ and
$\sigma^0(\textbf{r},\textbf{r}')$ are, of course, arbitrary,
whereas the final equilibrium mean and covariance, $n^{eq}
(\textbf{r})$ and $\sigma^{eq}(\textbf{r},\textbf{r}')$, are
univocally dictated by the external constraints imposed on the
system (isolation, contact with reservoirs, etc.) and by the
external field $\psi (\textbf{r})$, according to the second law of
thermodynamics. We open this section with a brief reference to the
specific thermodynamic framework in which this problem is embedded.
Given its conceptual importance, we provide additional details on
this topic in Appendix A. In the rest of this section we elaborate
the dynamic aspects as a concrete application of the generalized
Onsager theory just reviewed.

\subsection{Thermodynamics of fluids in spatially inhomogeneous states.}\label{VII.1}

The most fundamental thermodynamic ingredient in the application of
this general theory is the \emph{fundamental thermodynamic relation
(FTR)} $S=S[\textbf{a}]$, which assigns a value of the entropy at
any possible values of the state variables
$(a_1,a_2,...,a_M)=\textbf{a}$ \cite{callen}. In practice, however,
we only need the first and second derivatives of $S[\textbf{a}]$,
which define the intensive parameters $ F_j[ {\textbf a}] \equiv
k_B^{-1} \left(
\partial S[ {\textbf a}] /
\partial a_j \right)$ and the thermodynamic matrix
$\mathcal{E}_{ij}[{\textbf a }] \equiv -(
\partial F_i[{\textbf a }]/\partial a_j)$. To specify the variables
$(a_1,a_2,...,a_M)=\textbf{a}$ of our problem, let us first mentally
partition the volume $V$ in a number $C$ of smaller portions (or
{\it cells}), whose internal energy, particle number, and volume, we
denote by $E^{(r)}, N^{(r)}$ and $V^{(r)}$, respectively, with
$r=1,2,...,C.$ Then, the fundamental thermodynamic relation of this
system reads $S=S[{\bf E,N,V}]$, where {\bf E,\ N} and {\bf V} are
\emph{C}-dimensional vectors with components $E^{(r)},N^{(r)}$, and
$V^{(r)}$ $(r=1,2,...,C).$ For the sake of simplicity let us assume
that the volumes $V^{(r)}$ are all equal, $V^{(r)}=\Delta V=V/C$,
and remain fixed, so that only the variables [{\bf E,N}] are needed
to define a thermodynamic state. This corresponds to the selection
${\bf a}\equiv [{\bf E}, {\bf N}]$.

Just like in ordinary classical thermodynamics, under some
circumstances one may prefer to express the FTR not in terms of the
variables $[{\bf E}, {\bf N}]$, which involve the local internal
energy $E^{(r)}$, but in terms of the particle number profile
\textbf{N} and of some form of ``local temperature". Such
representation is most convenient under conditions in which the $N$
particle system is in contact with a thermal reservoir (in our case
the supporting solvent) that keeps temperature constant and uniform.
Under these circumstances the chemical equation of state can be
expresses as the dependence $\mu^{(r)}=\mu^{(r)}[\beta^R;{\bf N}]$
of the local electrochemical potential on the profile \textbf{N} and
of the thermal reservoir parameter $\beta^R$, as it is explained in
the appendix. In the general expression for $\mu^{(r)}[\beta^R;{\bf
N}]$ in Eq. (\ref{1}) below the explicit reference to the parameter
$\beta^R$ is omitted.

Let us notice that the discussion above is independent of the
spatial resolution employed to describe the non-uniformity of the
distribution of matter and energy, i.e., on the number $C$ of cells
in which we mentally partitioned the total volume $V$. Since this is
a mere informatic concept (the cells are not meant to represent
macroscopic subsystems), one can take the limit of maximum
resolution $\Delta V \to 0$ (or $C\to\infty$). Although no new
concepts arise in taking this limit, the notation and the
nomenclature change somewhat. For this, let us define the vectors
\textbf{n} and \textbf{e} whose components are the local particle
number density $n^{(r)}\equiv N^{(r)}/\Delta V$ and the local energy
density $e^{(r)}\equiv E^{(r)}/\Delta V$, and whose average remain
finite in this limit. Second, rather than labeling the cells with
the discrete index $r$, running from 1 to $C$, now we label them
with the position vector {\bf r} of their centers. In the limit of
vanishingly small cells, {\bf r} varies continuously in the volume
V, and hence, the vector components $n^{(r)}$ and $e^{(r)}$ with
$r=1,2,...,C$ now become the {\it functions} $n({\bf r})$ and
$e({\bf r})$ of the position vector ${\bf r}\in V$. As a
consequence, what used to be an ordinary function of the vectors
{\bf N} and {\bf E}, such as the entropy, now becomes what is called
a {\it functional} of the functions $n({\bf r})$ and $e({\bf r})$.
For example, the local electrochemical potential
$\mu^{(r)}=\mu^{(r)}[\beta^R;{\bf n}]$ of the particles at cell $r$
now becomes an ordinary function of the position vector {\bf r}, and
a functional of $n({\bf r})$. This dependence will be indicated as
$\mu[{\bf r};\beta^R,n]$ or simply as $\mu[\textbf{r};n]$. Of
course, the ordinary derivative of a function, such as the
thermodynamic matrix $\mathcal{E}^{(r,r')}[\textbf{N};\beta]\equiv
\left({\partial \beta \mu^{(r)}[\textbf{N};\beta] /
\partial N^{(r')}}\right)$ in Eq. (\ref{covariance}) of the appendix, now becomes
the functional derivative $\left({\delta \beta \mu[\textbf{r};n] /
\delta n(\textbf{r}')}\right)$. In the continuum limit, we must also
replace $(\Delta V)\sum_r$ by a volume integral $\int d^3r$  on the
vector {\bf r}, and $\delta_{r,r'}/\Delta V$ by the Dirac delta
function, $\delta({\bf r}-{\bf r}')$. Clearly, ``matrices" such as
$u^{(r,r')}$ now become functions of the two position vectors {\bf
r} and ${\bf r'}$, and matrix products now become convolutions.

With this notation, let us now write the most general expression for
the local electrochemical potential $\mu [{\bf r};n(t)] $  at
position ${\bf r}$ in units of the thermal energy $k_BT=\beta^{-1}$,
namely \cite{evans},

\begin{eqnarray}\label{1}
\beta\mu [{\bf r};n] & & =
\beta\mu^{in} [{\bf r};n] +  \beta \psi({\bf r}) \\
&& \equiv  \beta\mu^{*}(\beta) + \ln n({\bf r}) -c[{\bf r};n] +
\beta \psi({\bf r}).\nonumber
\end{eqnarray}
In this equation $\psi({\bf r})$ is the potential of the external
field acting on a particle at position \textbf{r}. The first two
terms of this definition of $\mu^{in} [{\bf r};n]$,
$(\beta\mu^{*}(\beta) + \ln n({\bf r}))$, are the ideal gas
contribution to the chemical potential, whereas the term $-c[{\bf
r};n]$ contains the deviations from ideal behavior due to
interparticle interactions.

Using Eq. (\ref{1}), the thermodynamic matrix $\mathcal{E}[{\bf
r},{\bf r}';n]\equiv \left[ {\delta \beta\mu [{\bf r};n]}/{\delta
n({\bf r}')}\right]$ can then be written in general as

\begin{equation}
\mathcal{E}[{\bf r},{\bf r}';n] = \delta({\bf r}-{\bf r}')/ n({\bf
r}) -c^{(2)}[{\bf r},{\bf r}';n], \label{stabmatrix}
\end{equation}
with $c^{(2)}[{\bf r},{\bf r}';n]\equiv (\delta c[{\bf r};n]/\delta
n({\bf r}'))$ being the \emph{functional }derivative of $c[{\bf
r};n]$ with respect to $n({\bf r}')$, referred to as the
\emph{direct correlation function}. On the other hand, the
covariance matrix $\sigma({\bf r},{\bf r}')= \overline{\delta n({\bf
r},0)\delta n({\bf r}',0)}$ can be written in terms of the
\emph{total correlation function} $h^{(2)}({\bf r},{\bf r}')$ as

\begin{equation}
\sigma({\bf r},{\bf r}') = n({\bf r}) \delta({\bf r}-{\bf r}') +
n({\bf r})n({\bf r}')h^{(2)}({\bf r},{\bf r}'). \label{covarmatrix}
\end{equation}
The matrices $\mathcal{E}[{\bf r},{\bf r}';n]$ and $\sigma({\bf
r},{\bf r}')$ are not in general related to each other. It is only
when they are evaluated at the equilibrium local concentration
profile $n^{eq}({\bf r})$ that they are related to each other by the
second equilibrium condition  in Eq. (\ref{covariance}) of the
appendix, which in the present notation reads

\begin{equation}\label{sigmaeqcond}
\int d\textbf{r}'\sigma^{eq}({\bf r},{\bf r}')\mathcal{E}[{\bf
r}',{\bf r}'';n^{eq}]=\delta(\textbf{r}-\textbf{r}'').
\end{equation}
Using Eqs. (\ref{stabmatrix}) and (\ref{covarmatrix}), one can
immediately see that this equation is equivalent to the well-known
Ornstein-Zernike equation \cite{mcquarrie}

\begin{equation}
h({\bf r},{\bf r}') = c({\bf r},{\bf r}') + \int d^3r'' c({\bf
r},{\bf r}'')n^{eq}({\bf r}'') h({\bf r}'',{\bf r}'), \label{oz}
\end{equation}
where $c({\bf r},{\bf r}')$ and $h({\bf r},{\bf r}')$ are,
respectively, the equilibrium value of $c^{(2)}({\bf r},{\bf r}')$
and $h^{(2)}({\bf r},{\bf r}')$.

In conclusion, the thermodynamic matrix $\mathcal{E}[{\bf r},{\bf
r}';n(t)]$ evaluated at an arbitrary state $n(t)$ is fully
determined by the chemical equation of state. Its equilibrium value,
$\mathcal{E}[{\bf r},{\bf r}';n^{eq}]$, determines the covariance
$\sigma^{eq}({\bf r},{\bf r}')$ of the equilibrium distribution by
means of Eq. (\ref{sigmaeqcond}), which is equivalent to the
Ornstein-Zernike equation above. The time-dependent covariance
$\sigma({\bf r},{\bf r}';t)$  of an arbitrary non-equilibrium state,
however, cannot be determined in this manner, unless the local
equilibrium approximation (i.e., the quasi-static limit) is assumed
to be valid. Thus, in general we need an independent,
non-thermodynamic condition, to determine this important property,
and this is the main subject of the following subsection.

\subsection{(Irreversible) Time evolution of $\overline{n}(\textbf{r},t)$
and $\sigma(\textbf{r},\textbf{r}';t)$.}

Let us start by writing the analog of  Eq. (\ref{releq0}). The
macroscopic diffusive relaxation of the local concentration
$\overline{n}(\textbf{r},t)$ of colloidal particles is described by
the most general non-linear but spatially and temporally local
diffusion equation provided by Fick's law, which reads
\cite{keizer,degrootmazur}
\begin{equation} \frac{\partial \overline{n}(\textbf{r},t)}{\partial
t} = D^0{\nabla} \cdot b(\textbf{r},t)\overline{n}(\textbf{r},t)
\nabla \beta\mu[{\bf r};\overline{n}(t)]. \label{difeqdl}
\end{equation}
In this equation $D_0$ is the diffusion coefficient of the colloidal
particles in the absence of interactions between them and
$b(\textbf{r},t)$ is a local reduced mobility, to be specified
later, which describes the frictional effects of the direct (i.e.,
conservative) interactions between particles, as deviations from the
value $b(\textbf{r},t)=1$.

We may now linearize this equation around the equilibrium profile
$n^{eq}({\bf r})$, to get the analog of Eq.
(\ref{releq0linearized}), from which we can identify the ``matrix"
$\mathcal{L} [\textbf{r},\textbf{r}';{\overline n}(t)]$ of Onsager
kinetic coefficients as

\begin{equation}
-\mathcal{L} [\textbf{r},\textbf{r}';{\overline  n }(t)] =
D^0{\nabla} \cdot \overline{n}(\textbf{r},t)  \
b(\textbf{r},t)\nabla \delta(\textbf{r}-\textbf{r}').
\label{matrixl3}
\end{equation}
Using Eq. (\ref{sigmadtirrev2}) we can then write the relaxation
equation for $\sigma(\textbf{r},\textbf{r}';t)$ as
\begin{eqnarray}
\begin{split}
\frac{\partial \sigma(\textbf{r},\textbf{r}';t)}{\partial t} = &
D^0{\nabla} \cdot \overline{n}(\textbf{r},t) \ b(\textbf{r},t)\nabla
\int d \textbf{r}_2
\mathcal{E}[\textbf{r},\textbf{r}_2;\overline{n}(t)]
\sigma(\textbf{r}_2,\textbf{r}';t) \\ & +  D^0{\nabla}' \cdot
\overline{n}(\textbf{r}',t) \ b(\textbf{r}',t)\nabla' \int d
\textbf{r}_2 \mathcal{E}[\textbf{r}',\textbf{r}_2;\overline{n}(t)]
\sigma(\textbf{r}_2,\textbf{r};t) \\ & -2D^0{\nabla} \cdot
\overline{n}(\textbf{r},t)  \ b(\textbf{r},t)\nabla
\delta(\textbf{r}-\textbf{r}'). \label{relsigmadif2}
\end{split}
\end{eqnarray}

Let us now describe the fluctuations $\delta n(\textbf{r},t+\tau)
\equiv n(\textbf{r},t+\tau)- \overline{n}(\textbf{r},t)$ of the
local concentration at position \textbf{r} and time $t+\tau$ around
the mean value $\overline{n}(\textbf{r},t)$ within a microscopic
temporal resolution described by the time $\tau$. The assumption of
local stationarity means that in the time-scale of $\tau$,
$\overline{n}(\textbf{r},t)$ is to be treated as a constant. We may
add the spatial counterpart of this simplifying assumption. Thus, we
write the fluctuations as $\delta n(\textbf{r}+\textbf{x},t+\tau)
\equiv n(\textbf{r}+\textbf{x},t+\tau)- \overline{n}(\textbf{r},t)$,
where the argument \textbf{r} of $\overline{n}(\textbf{r},t)$ refer
to the macroscopic resolution of the measured variations of the
local equilibrium profile, whereas the position vector \textbf{x}
adds the possibility of microscopic resolution in the description of
the thermal fluctuations. Defining the fluctuations as the
deviations of the microscopic local concentration profile
$n(\textbf{r}+\textbf{x},t+\tau)$ from the mean value
$\overline{n}(\textbf{r},t)$ indicates that, within the microscopic
spatial variations described by the position vector \textbf{x},
$\overline{n}(\textbf{r},t)$ must be treated as a constant. To a
large extent, this is equivalent to recover the partitioning of the
system in cells of a small but finite volume $\Delta V$, and assume
that variations from cell to cell are described by the vector
\textbf{r}, whereas variations within cells are described by the
vector \textbf{x}, and that within the intra-cell scale, the system
can be regarded as uniform and isotropic. Under these conditions,
the covariance
$\sigma(\textbf{r}+\textbf{x},\textbf{r}+\textbf{x}';t)$ may be
written as $\sigma(\mid\textbf{x}-\textbf{x}'\mid; \textbf{r},t)$,
and in terms of its Fourier transform $\sigma(k;\textbf{r},t)$, as

\begin{equation}
\sigma(\mid\textbf{x}-\textbf{x}'\mid;\textbf{r},t)=
\frac{1}{(2\pi)^3}\int d^3 k e^{-i\textbf{k}\cdot
(\textbf{x}-\textbf{x}')} \sigma(k;\textbf{r},t). \label{ftsigma}
\end{equation}
In this manner, Eq. (\ref{relsigmadif2}) may be re-written as
\begin{eqnarray}
\begin{split}
\frac{\partial \sigma(k;\textbf{r},t)}{\partial t} = & -2k^2 D^0
\overline{n}(\textbf{r},t) b(\textbf{r},t)
\mathcal{E}(k;\overline{n}(\textbf{r},t)) \sigma(k;\textbf{r},t)
\\ & +2k^2 D^0 \overline{n}(\textbf{r},t)\ b(\textbf{r},t), \label{relsigmadif2p}
\end{split}
\end{eqnarray}
where $\mathcal{E}(k;\overline{n}(\textbf{r},t))$ is the FT of
$\mathcal{E}(\mid\textbf{x}-\textbf{x}'\mid;
\overline{n}(\textbf{r},t))$, defined as the thermodynamic matrix
evaluated at a uniform concentration profile with a constant value
given by the local and instantaneous concentration
$\overline{n}(\textbf{r},t)$ at position \textbf{r} and time
\emph{t}.

\subsection{Relaxation equation for $C(t,t')$}

The description of the fluctuations $\delta
n(\textbf{r}+\textbf{x},t+\tau) \equiv
n(\textbf{r}+\textbf{x},t+\tau)- \overline{n}(\textbf{r},t)$ with
the temporal and spatial resolution described by the time $\tau$ and
position vector \textbf{x} cannot be obtained by simply linearizing
the macroscopic version of Fick's diffusion equation above. Instead,
one has to consider a generalized version of Fick's law, which
contains Eq. (\ref{difeqdl}) as its macroscopic limit. Such an
extension reads
\begin{equation}
\frac{\partial n(\textbf{r},t)}{\partial t} = D^0{\nabla} \cdot
\int_0^t dt'\int d^3r' \
b[\textbf{r}-\textbf{r}';t-t']n(\textbf{r}',t')\nabla'\beta\mu[{\bf
r}';n(t')],  \label{relaxation3}
\end{equation}
where $b[\textbf{r};t]$ is a time-dependent local mobility that must
yet be specified.

This generalized diffusion equation may be derived rather simply by
complementing the continuity equation,
\begin{equation}
\frac{\partial n(\textbf{r},t)}{\partial t} = - {\nabla} \cdot
\textbf{j} (\textbf{r},t), \label{continuity}
\end{equation}
with a constitutive relation constructed at the level of the
particle current. We require that the friction force on the
particles in the neighborhood of position \textbf{r} must be
equilibrated by the osmotic force $-\nabla\mu^{in} [{\bf r};n] $ and
by the external force $-\nabla\Psi ({\bf r})$ on each particle, both
included in $-\nabla\mu [{\bf r};n]$, so that

\begin{equation}
 \zeta ^0 \textbf{j} (\textbf{r},t)+\int_0^t dt'\int d^3r' \ \Delta
\zeta [\textbf{r}-\textbf{r}';t-t']\cdot \textbf{j}
(\textbf{r},t)=-n(\textbf{r},t)\nabla\mu[{\bf r};n(t)],
\label{continuity}
\end{equation}
The friction force per unit volume on the left hand side of this
equation is the sum of the friction due to the supporting solvent,
$\zeta^0 \textbf{j} (\textbf{r},t)$, and the frictional effects due
to the interactions between the colloidal particles themselves,
$(\Delta\zeta) \textbf{j} (\textbf{r},t)$. The latter, however, is
assumed to be in general spatially and temporally nonlocal. The
solution of this equation for $\textbf{j} (\textbf{r},t)$ can be
written as
\begin{equation}\label{udrt2}
\textbf{j} (\textbf{r},t) =-D^0\int_0^t dt' \int d^3r'
b[\textbf{r}-\textbf{r}';t-t']n(\textbf{r}',t')\nabla' \beta\mu[{\bf
r}';n(t')],
\end{equation}
where $D^0$ is the free diffusion coefficient, defined here as
$D^0\equiv k_BT/\zeta^0$, and where the spatially  and temporally
non-local mobility kernel $b[\textbf{r}-\textbf{r}';t]$ is defined
in terms of the memory function $\Delta \zeta^*
[\textbf{r}-\textbf{r}';t-t']\equiv \Delta \zeta
[\textbf{r}-\textbf{r}';t-t']/\zeta^0$ as the solution of the
equation

\begin{equation}\label{jdrt}
\textbf{b} [\textbf{r}-\textbf{r}';t] =
\delta(\textbf{r}-\textbf{r}')2\delta(t)-\int_0^t dt' \int d^3r''
\Delta \zeta^*
[\textbf{r}-\textbf{r}'';t-t']b[\textbf{r}''-\textbf{r}';t'].
\end{equation}
Using Eq. (\ref{udrt2}) in the continuity equation
(\ref{continuity}) finally leads us to Eq. (\ref{relaxation3}),
which reduces to Eq. (\ref{difeqdl}) when the generalized mobility
kernel $b[\textbf{r}-\textbf{r}';t-t']$ is approximated by its
spatially and temporally local limit,
\begin{equation}
b[\textbf{r}-\textbf{r}';t-t']=b(\textbf{r},t)\delta(\textbf{r}-\textbf{r}')2
\delta(t-t'), \label{label1}
\end{equation}
where
\begin{equation}
b(\textbf{r},t)\equiv \int d\textbf{x}\int_0^\infty d\tau \
b[\textbf{x},\tau;\textbf{r},t] \label{bast}
\end{equation}
with $b[\textbf{x},\tau;\textbf{r},t]\equiv
b[(\textbf{r}+\textbf{x})-\textbf{r};(t+\tau)-t]$.

We can now proceed to identify the elements of Eq.
(\ref{fluctuations1}) corresponding to our problem. In the present
case, the corresponding antisymmetric matrix
$\omega[\overline{\textbf{a}}(t)]$ vanishes due to time-reversal
symmetry arguments \cite{delrio}. We can then write the matrix
$\gamma[\tau;\overline{\textbf{a}}(t)]$ as the non-markovian and
spatially non-local Onsager matrix implied by the general diffusion
equation in Eq. (\ref{relaxation3}), which must reflect, in
addition, that within the temporal and spatial resolution of the
variables \textbf{x} and $\tau$, the local concentration profile
$\overline{n}(\textbf{r},t)$ remains uniform and stationary. These
assumptions can be summarized by the following stochastic equation
for $\delta n(\textbf{r}+\textbf{x},t+\tau)$

\begin{eqnarray}
\begin{split}\label{fluct}
\frac{\partial \delta n(\textbf{r}+\textbf{x},t+\tau)}{\partial
\tau} =   & D^0\overline{n}(\textbf{r},t){\nabla}_\textbf{x} \cdot
\int_0^\tau d\tau'\int d\textbf{x}_1
b[\textbf{x}-\textbf{x}_1,\tau-\tau';\textbf{r},t]\nabla_{\textbf{x}_1}  \\
& \int d\textbf{x}_2
 \sigma^{-1}(\mid\textbf{x}_1-\textbf{x}_2\mid;t)\delta
n(\textbf{r}+\textbf{x}_2,t+\tau')  +
\textbf{f}(\textbf{r}+\textbf{x},t+\tau),
\end{split}
\end{eqnarray}
where the function $ \sigma^{-1}(\mid\textbf{x}-\textbf{x}'\mid;t)$
is the inverse of the covariance $
\sigma(\mid\textbf{x}-\textbf{x}'\mid;t)$ (in the sense that their
convolution equals the Dirac delta function), so that its Fourier
transform is $1/\sigma(k;\textbf{r},t)$. The random term
$\textbf{f}(\textbf{r}+\textbf{x},t+\tau)$ of eq. (\ref{fluct}) is
assumed to have zero mean and time correlation function given by
$<\textbf{f}(\textbf{r}+\textbf{x},t+\tau)\textbf{f}^{\dagger}
(\textbf{r}+\textbf{x}',t+\tau')>=\gamma[\textbf{x}-\textbf{x}',\tau-\tau';\textbf{r},t]$,
with

\begin{equation}\label{nonmarkovmatrixl}
\gamma[\textbf{x}-\textbf{x}',\tau;\textbf{r},t] \equiv
D^0\overline{n}(\textbf{r},t){\nabla}_\textbf{x} \cdot \int
d\textbf{x}_1 b[\textbf{x}-\textbf{x}_1,\tau;
\textbf{r},t]\nabla_{\textbf{x}_1} \delta(\textbf{x}_1-\textbf{x}').
\end{equation}

Similarly, the analog  of Eq. (\ref{fluctuations3}) for the time
correlation function $C(\tau,t)$ is the relaxation equation for
$C(\textbf{x},\tau;\textbf{r},t) \equiv \overline{\delta
n(\textbf{r}+\textbf{x},t+\tau)\delta n(\textbf{r},t)}$, namely,
\begin{eqnarray}
\begin{split}
\frac{\partial C(\textbf{x},\tau;\textbf{r},t) }{\partial \tau} = &
D^0\overline{n}(\textbf{r},t){\nabla}_\textbf{x} \cdot \int_0^\tau
d\tau'\int d\textbf{x}_1 b[\textbf{x}-\textbf{x}_1,\tau-\tau';
\textbf{r},t]\nabla_{\textbf{x}_1} \\ & \int d\textbf{x}_2
\sigma^{-1}(\textbf{x}_1,\textbf{x}_2;t)C(\textbf{x}_2,
\tau';\textbf{r},t). \label{ctdtau2}
\end{split}
\end{eqnarray}

\subsection{Approximate self-consistent closure for the local
mobility $b(\textbf{r},t)$}\label{IV.4}

The generalized theory of non-equilibrium diffusion just presented
writes the relaxation of the mean value
$\overline{n}(\textbf{r},t)$, of the covariance
$\sigma(\textbf{r},\textbf{r}';t)$, and of the two-time correlation
function $C(\textbf{x},\tau; \textbf{r},t)$, through Eqs.
(\ref{difeqdl}), (\ref{relsigmadif2}) (or (\ref{relsigmadif2p})),
and (\ref{ctdtau2}), in terms of the generalized mobility
$b[\textbf{x},\tau; \textbf{r},t]$ or, according to Eq.
(\ref{jdrt}), in terms of the temporally and spatially nonlocal
friction function $\Delta \zeta [\textbf{x},\tau; \textbf{r},t]$.
These equations constitute the general framework in which
approximations may be introduced to construct a closed system of
equations for the properties involved. The main purpose of the
present subsection is to determine an independent closure relation
for the local mobility $b(\textbf{r};t)$,  needed in Eqs.
(\ref{difeqdl}) and (\ref{relsigmadif2}) (or (\ref{relsigmadif2p})),
in terms of $\overline{n}(\textbf{r};t)$and
$\sigma(\textbf{r},\textbf{r}';t)$.

This, however, will be a relatively involved process. The reason for
this is that, according to Eq. (\ref{bast}), the local mobility
$b(\textbf{r};t)$ is an integral of the non-local generalized
mobility $b[\textbf{x},\tau;\textbf{r},t]$ appearing in Eq.
(\ref{ctdtau2}) for the time-correlation function
$C(\textbf{x},\tau; \textbf{r},t)$. Thus, the determination of
$b[\textbf{x},\tau;\textbf{r},t]$, is essentially equivalent to the
determination of $C(\textbf{x},\tau; \textbf{r},t)$, which is
intrinsically an involved and rich problem, even under ordinary
equilibrium conditions. Thus, our answer to this problem is
equivalent to extending to non-equilibrium conditions the
equilibrium theoretical approach to calculate these dynamic
properties.

With this aim let us refer to Eq. (\ref{ctdtau2}) and assume that,
within the approximation of local uniformity and isotropy introduced
above, $C(\textbf{x},\tau;\textbf{r},t)=
C(\mid\textbf{x}\mid,\tau;\textbf{r},t)$. We then write the Fourier
transform (FT) of this correlation function as
\begin{equation}
C(\mid\textbf{x}\mid,\tau;\textbf{r},t)=
\frac{1}{(2\pi)^3}\int d^3 k
e^{-i\textbf{k}\cdot \textbf{x}} C(k,\tau;\textbf{r},t),
\end{equation}

Denoting also the FT of $b[\mid\textbf{x}\mid,\tau; \textbf{r},t]$
as $b(k,\tau; \textbf{r},t)$, we can then rewrite Eq.
(\ref{ctdtau2}) in Fourier space as
\begin{equation}
\frac{\partial C(k,\tau;\textbf{r},t) }{\partial \tau} = -k^2 D^0
\overline{n}(\textbf{r},t)\int_0^\tau d\tau'\int b(k,\tau-\tau';
\textbf{r},t) \sigma^{-1}(k;\textbf{r},t)C(k, \tau';\textbf{r},t).
\label{ctdtau2ft}
\end{equation}

In its turn, the mobility $b(k,\tau; \textbf{r},t)$ can be expressed
in terms of the FT $\Delta \zeta(k,\tau; \textbf{r},t)$ of $\Delta
\zeta(\mid\textbf{x}\mid,\tau; \textbf{r},t)$ according to Eq.
(\ref{jdrt}), which in Laplace space reads

\begin{equation}
\hat b(k,z; \textbf{r},t)= \left[1+\Delta\hat {\zeta}^*(k,z;
\textbf{r},t)\right]^{-1} \label{bdkz}
\end{equation}
with $\Delta\hat {\zeta}^*(k,z; \textbf{r},t)\equiv \Delta\hat
{\zeta}(k,z; \textbf{r},t)/\zeta^0$ and with the hat and the
argument $z$ meaning Laplace transform (LT). Using this result in
the Laplace-transformed version of eq. (\ref{ctdtau2ft}), we finally
get the following expression for the LT of $C(k,\tau;\textbf{r},t)$
in terms of $\Delta \zeta^*(k,z; \textbf{r},t)$

\begin{gather}\label{fluct5}
\hat C(k,z; \textbf{r},t) = \frac{\sigma(k; \textbf{r},t)}{z+\frac{k^2D^0
\overline{n}(\textbf{r},t)\sigma^{-1}(k; \textbf{r},t)}
{1+\Delta \hat \zeta^*(k,z; \textbf{r},t)}}.
\end{gather}

Let us notice that we can also introduce the notation by
$C(k,\tau;\textbf{r},t)=\overline{n}(\textbf{r},t)F(k,\tau;\textbf{r},t)$,
with $F(k,\tau;\textbf{r},t)$ being the non-equilibrium intermediate
scattering function, whose initial value $ F(k,\tau=0;\textbf{r},t)=
S(k;\textbf{r},t)$ defines the time-evolving spatially local static
structure factor $S(k;\textbf{r},t)$. With this more familiar
notation it is not difficult to recognize in Eq. (\ref{fluct5}) the
non-equilibrium extension of the well-known exact expression for the
LT of the intermediate scattering function in terms of the so-called
irreducible memory function $\Delta\hat \zeta^*(k,z; \textbf{r},t)$
\cite{1,2,scgle0}. There is, however, a deep fundamental difference
between this expression for $\hat C(k,z; \textbf{r},t)$ and its
equilibrium counterpart: the initial value $\sigma(k;
\textbf{r},t)=\overline{n}(\textbf{r};t)S(k; \textbf{r},t)$ needed
in Eq. (\ref{fluct5}) derives from the nonequilibrium solution  of
the relaxation equation in Eq. (\ref{relsigmadif2p}), and \emph{not}
from the local equilibrium approximation
$\sigma^{l.e.}(k;\overline{n}(\textbf{r},t))=
[\mathcal{E}(k;\overline{n}(\textbf{r},t))]^{-1}$. Of course, the
general expression in Eq. (\ref{fluct5}) contains the conventional
equilibrium result as the particular case in which the static
structure factor $S(k; \textbf{r},t)=\sigma(k;
\textbf{r},t)/\overline{n}(\textbf{r};t)$ is given by its
equilibrium value $S^{eq}(k;\overline{n}^{eq})= [
\overline{n}^{eq}\mathcal{E}(k;\overline{n}^{eq})]^{-1}$.

Let us mention that the equilibrium counterpart of Eq.
(\ref{fluct5}) can also be derived without appealing to the
phenomenological non-linear and non-local extension of Fick's
diffusion equation in Eq. (\ref{relaxation3}). Thus, in Ref.
\cite{faraday} the non-Markovian extension of Onsager's theory
(referred to there as the ``generalized Langevin equation" (GLE)
approach) was employed to derive the equilibrium version of Eq.
(\ref{fluct}), from which the equilibrium version of Eq.
(\ref{fluct5}) follows. The value of the phenomenological derivation
of the non-linear Fick's diffusion equation of Eq.
(\ref{relaxation3}) is that it is a natural non-linear extension of
the more rigorously-derived equilibrium linear theory. A similar
situation arises when one considers the derivation of the result
analogous to Eq. (\ref{fluct5}) for the self component $\hat
C_S(k,z; \textbf{r},t)$ of $\hat C(k,z; \textbf{r},t)$. This result
that can also be derived in either of these two manners, both of
which lead to the following expression for  $\hat C_S(k,z;
\textbf{r},t)$

\begin{gather}\label{fluct5s}
\hat C_S(k,z; \textbf{r},t) = \frac{1}{z+\frac{k^2D^0 }{1+\Delta
\hat \zeta^*_S(k,z; \textbf{r},t)}}.
\end{gather}

In this manner, Eqs. (\ref{fluct5}) and (\ref{fluct5s}) write the
non-equlibrium collective and self time-correlation functions $\hat
C(k,z; \textbf{r},t)$ and $\hat C_S(k,z; \textbf{r},t)$ in terms of
the respective irreducible memory functions $\Delta \hat\zeta^*(k,z;
\textbf{r},t)$ and $\Delta \hat\zeta^*_S(k,z; \textbf{r},t)$. At
this point, with the aim of establishing a self-consistent scheme
for the calculation of these four properties, we propose to proceed
along the same lines, and to adopt the same approximations, of the
equilibrium SCGLE theory in its simplest formulation \cite{todos2}.
Thus, we start by adopting the Vineyard approximation

\begin{equation}
\Delta \hat\zeta^*(k,z; \textbf{r},t)= \Delta \hat\zeta^*_S(k,z;
\textbf{r},t), \label{vineyard0}
\end{equation}
along with the factorization approximation

\begin{equation}
\Delta \hat\zeta^*(k,z; \textbf{r},t)= \lambda (k; \textbf{r},t)\
\Delta \hat\zeta^*(z; \textbf{r},t), \label{interpolation}
\end{equation}
in which the function $\lambda (k; \textbf{r},t)$ is a
phenomenological ``interpolating function" \cite{todos1,todos2},
given by

\begin{equation}
\lambda (k; \textbf{r},t)=\frac{1}{1+\left( \frac{k}{k_{c}}\right)
^{2}}, \label{lambdadk}
\end{equation}
where $k_{c } \gtrsim 2\pi/d$, where $d$ is some form of distance of
closest approach. A simple empirical prescription is to choose $k_{c
}$ as $k_{c }= k_{\min }$, the position of the first minimum (beyond
the main peak) of the non-equilibrium static structure factor $S(k;
\textbf{r},t)=\sigma(k; \textbf{r},t)/\overline{n}(\textbf{r};t)$ at
position \textbf{r} and time \emph{t}.

The function $\Delta \hat\zeta^* (z; \textbf{r},t)$ in Eq.
(\ref{interpolation}) is the Laplace transform of the
$\tau$-dependent friction function $\Delta \zeta^* (\tau;
\textbf{r},t)\equiv \Delta \zeta (\tau; \textbf{r},t)/\zeta_0$,
which can be approximated by the following expression
\begin{equation}
\Delta \zeta^* (\tau; \textbf{r},t)=\frac{D_0}{3\left( 2\pi \right)
^{3}\overline{n}(\textbf{r},t)}\int d {\bf k}\ k^2 \left[\frac{ S(k;
\textbf{r},t)-1}{S(k; \textbf{r},t)}\right]^2 F(k,\tau;
\textbf{r},t)F_S(k,\tau; \textbf{r},t). \label{dzdt}
\end{equation}
The derivation of this expression follows, in a first approximation,
essentially the same arguments employed in the derivation of its
equilibrium counterpart, explained in the original presentation in
Ref. \cite{faraday} (also reviewed in appendix B of Ref.
\cite{todos1}). The main aspect that needs to be adapted refers to
the statistical distribution of the local concentration profile of
the particles around a particular tracer particle, whose mean and
covariance in the original derivation refers to the equilibrium
distribution, whereas now they refer to the mean and covariance of
the statistical distribution representing a non-equilibrium state.
In this manner, the exact results in Eqs.\ (\ref{fluct5}) and
(\ref{fluct5s}), complemented with the closure relation for the
time-dependent friction function in Eq.\ (\ref{dzdt}) and the
Vineyard and the factorization approximations in Eqs.\
(\ref{vineyard0}), (\ref{interpolation}) and (\ref{lambdadk}),
constitute a closed system of equations that must be solved
self-consistently.

\section{Full non-equilibrium theory and particular limits}\label{V}

In summary, the NE-SCGLE theory is defined in terms of a system of
equations for the time-evolution of the mean value
$\overline{n}(\textbf{r},t)$ and of the covariance
$\sigma(\textbf{r},\textbf{r}';t)$ of the fluctuations of the local
concentration profile $n(\textbf{r},t)$ of colloidal particles,
namely,
\begin{equation} \frac{\partial \overline{n}(\textbf{r},t)}{\partial
t} = D^0{\nabla} \cdot b(\textbf{r},t)\overline{n}(\textbf{r},t)
\nabla \beta\mu[{\bf r};\overline{n}(t)] \label{difeqdlp}
\end{equation}
and
\begin{eqnarray}
\begin{split}
\frac{\partial \sigma(\textbf{r},\textbf{r}';t)}{\partial t} = &
D^0{\nabla} \cdot \overline{n}(\textbf{r},t) \ b(\textbf{r},t)\nabla
\int d \textbf{r}_2
\mathcal{E}[\textbf{r},\textbf{r}_2;\overline{n}(t)]
\sigma(\textbf{r}_2,\textbf{r}';t) \\ & +  D^0{\nabla}' \cdot
\overline{n}(\textbf{r}',t) \ b(\textbf{r}',t)\nabla' \int d
\textbf{r}_2 \mathcal{E}[\textbf{r}',\textbf{r}_2;\overline{n}(t)]
\sigma(\textbf{r}_2,\textbf{r};t) \\ & -2D^0{\nabla} \cdot
\overline{n}(\textbf{r},t)  \ b(\textbf{r},t)\nabla
\delta(\textbf{r}-\textbf{r}'). \label{relsigmadif2p}
\end{split}
\end{eqnarray}
with $\mathcal{E}[{\bf r},{\bf r}';n]\equiv \left[ {\delta \beta\mu
[{\bf r};n]}/{\delta n({\bf r}')}\right]$. We assume that we can
approximate this thermodynamic matrix as
$\mathcal{E}[\textbf{r},\textbf{r}+\textbf{x};\overline{n}(t)]\approx
\mathcal{E}(|\textbf{x}|;\overline{n}(\textbf{r},t))$, i.e., by the
thermodynamic matrix evaluated at a uniform concentration profile
with a constant value given by the local and instantaneous
concentration $\overline{n}(\textbf{r},t)$ at position \textbf{r}
and time \emph{t}. Then the covariance can also be approximated as
$\sigma(\textbf{r},\textbf{r}+\textbf{x};t)\approx
\sigma(|\textbf{x}|;\textbf{r},t)$, and the latter equation can also
be written as
\begin{eqnarray}
\begin{split}
\frac{\partial \sigma(k;\textbf{r},t)}{\partial t} = & -2k^2 D^0
\overline{n}(\textbf{r},t) b(\textbf{r},t)
\mathcal{E}(k;\overline{n}(\textbf{r},t)) \sigma(k;\textbf{r},t)
\\ & +2k^2 D^0 \overline{n}(\textbf{r},t)\ b(\textbf{r},t), \label{relsigmadif2p}
\end{split}
\end{eqnarray}
where $\sigma(k;\textbf{r},t) \equiv \int d^3 k e^{i\textbf{k}\cdot
\textbf{x}} \sigma(|\textbf{x}|;\textbf{r},t)$ and
$\mathcal{E}(k;\overline{n}(\textbf{r},t)) \equiv (2\pi)^{-3}\int
d^3 k e^{-i\textbf{k}\cdot \textbf{x}}
\mathcal{E}(|\textbf{x}|;\overline{n}(\textbf{r},t))$.

Besides the chemical equation of state (i.e., the functional
dependence of the local electrochemical potential $\beta\mu[{\bf
r};\overline{n}(t)]$ on the concentration profile
$n(\textbf{r},t)$), the solution of these equations require the
simultaneous determination of the local mobility function
$b(\textbf{r},t)$ which is given, according to Eqs. (\ref{bast}),
(\ref{bdkz}), and (\ref{interpolation}), by
\begin{equation}
b(\textbf{r},t)= \left[1+\int_0^\infty d\tau \Delta \zeta^*(\tau;
\textbf{r},t)\right]^{-1}. \label{bdkzfinal}
\end{equation}
The actual calculation of $b(\textbf{r},t)$ requires the solution,
at each position \textbf{r} and each evolution time $t$, of a system
of equations involving the Laplace transform (LT) $\hat C(k,z;
\textbf{r},t) \equiv \int_0^\infty d \tau C(k,\tau; \textbf{r},t)$
of $ C(k,\tau; \textbf{r},t)$ and of its \emph{self} component
$C_S(k,\tau; \textbf{r},t)$, as well as the LT of the
$\tau$-dependent friction function $\Delta \zeta^* (\tau;
\textbf{r},t)$, namely,
\begin{gather}\label{fluct5p}
\hat C(k,z; \textbf{r},t) = \frac{\sigma(k;
\textbf{r},t)}{z+\frac{k^2D^0 \overline{n}(\textbf{r},t)
\sigma^{-1}(k; \textbf{r},t)}{1+\lambda (k; \textbf{r},t)\ \Delta
\hat\zeta^*(z; \textbf{r},t)}},
\end{gather}

\begin{gather}\label{fluct5sp}
\hat C_S(k,z; \textbf{r},t) = \frac{1}{z+\frac{k^2D^0 }{1+\lambda
(k; \textbf{r},t)\ \Delta \hat\zeta^*(z; \textbf{r},t)}},
\end{gather}
and
\begin{equation}
\Delta \zeta^* (\tau; \textbf{r},t)=\frac{D_0}{3\left( 2\pi \right)
^{3}}\int d {\bf k}\ k^2 \left[\frac{ \sigma(k;
\textbf{r},t)/\overline{n}(\textbf{r},t)-1}{\sigma(k;
\textbf{r},t)}\right]^2 C(k,\tau; \textbf{r},t)C_S(k,\tau;
\textbf{r},t), \label{dzdtp}
\end{equation}
with $\lambda (k; \textbf{r},t)$ being the phenomenological
``interpolating function" given by Eq. (\ref{lambdadk}).

An important aspect of Eqs. (\ref{fluct5p})-(\ref{dzdtp}) above
refers to their long-$\tau$ (or small $z$) asymptotic stationary
solutions, referred to as the non-ergodicity parameters of the
corresponding dynamic properties. These are given by
\begin{equation}
f(k;\textbf{r},t)\equiv \lim_{\tau\to\infty}
\frac{C(k,\tau;\textbf{r},t)}{\sigma(k;\textbf{r},t)} = \frac
{\lambda(k;\textbf{r},t)\sigma(k;\textbf{r},t)}{\lambda(k;\textbf{r},t)
\sigma(k;\textbf{r},t)+k^2\overline{n}(\textbf{r},t)\gamma(\textbf{r},t)}
\label{fdkinf}
\end{equation}
and

\begin{equation}
f_S(k;\textbf{r},t)\equiv \lim_{\tau\to\infty}
C_S(k,\tau;\textbf{r},t) = \frac
{\lambda(k;\textbf{r},t)}{\lambda(k;\textbf{r},t)
+k^2\gamma(\textbf{r},t)}, \label{fdksinf}
\end{equation}
where the (spatially and temporally dependent) squared localization
length $\gamma (\textbf{r},t)$ is the solution of
\begin{equation}
\frac{1}{\gamma(\textbf{r},t)} = \frac{1}{6\pi^{2}}\int_{0}^{\infty
} dkk^4\frac{\left[\sigma(k;
\textbf{r},t)/\overline{n}(\textbf{r},t)-1\right] ^{2}\lambda^2
(k;\textbf{r},t)}{\left[\lambda
(k;\textbf{r},t)\sigma(k;\textbf{r},t) +
k^2\overline{n}(\textbf{r},t)\gamma(\textbf{r},t)\right]\left[\lambda
(k;\textbf{r},t) + k^2\gamma(\textbf{r},t)\right]}. \label{nep5pp}
\end{equation}
If the solution ${\gamma(\textbf{r},t)}$ of the latter equation is
infinite, we can say that at that position \textbf{r} and waiting
time $t$ the system still remains ergodic, but if it is finite, we
say that the system became dynamically arrested.

\subsection{Particular cases and limits}\label{IV.6}

Eqs. (\ref{difeqdlp})-(\ref{nep5pp}) constitute the full
non-equilibrium SCGLE theory. Let us recall that
$C(k,\tau;\textbf{r},t)=
\overline{n}(\textbf{r},t)F(k,\tau;\textbf{r},t)$, with
$F(k,\tau;\textbf{r},t)$ being the non-equilibrium intermediate
scattering function, whose initial value $ F(k,\tau=0;\textbf{r},t)=
S(k;\textbf{r},t)$ is the time-evolving spatially varying static
structure factor $S(k;\textbf{r},t)= \sigma(k;\textbf{r},t)/
\overline{n}(\textbf{r},t)$. With this notation, Eqs.
(\ref{difeqdlp})-(\ref{nep5pp}) will probably appear more familiar.
In fact, it is not difficult to recognize in these general equations
a number of relevant concepts when adequate limits or cases are
considered, some of which are discussed in what follows.

The first obvious general limit to discuss refers to the long
evolution-time limit, $t \to \infty$. Assuming static external
fields and static thermodynamic constraints, one expects that in
this limit the solution of Eqs. (\ref{difeqdlp}) and
(\ref{relsigmadif2p}) will converge to a stationary state, denoted
by $\overline{n}^{ss}(\textbf{r})$ and $\sigma^{ss}(k;\textbf{r})$.
This stationary state will be a thermodynamic equilibrium state if
$(\partial \overline{n}^{ss}(\textbf{r})/\partial t)$ and $(\partial
\sigma^{ss}(k;\textbf{r})/\partial t)$ vanish due to the fact that
the two equilibrium conditions, $\nabla
\mu[\textbf{r};\bar{n}^{ss}]=0$ and
$\mathcal{E}(k;\overline{n}^{ss}(\textbf{r}))
\sigma^{ss}(k;\textbf{r})=1$, have been attained. Other stationary
solutions of Eqs. (\ref{difeqdlp})-(\ref{relsigmadif2p}) might,
however, exist in which the derivatives $(\partial
\overline{n}^{ss}(\textbf{r})/\partial t)$ and $(\partial
\sigma^{ss}(k;\textbf{r})/\partial t)$ vanish due to vanishing of
the local mobility $b(\textbf{r},t\to \infty)$, a condition for
dynamic arrest. We may, however, disregard the consequences of this
second possibility, and assume that the system will always be able
to reach its thermodynamic equilibrium state. Furthermore, let us
assume that the system is not subjected to external fields, so that
$ \overline{n}^{ss}(\textbf{r})= \overline{n}_b$ and
$\sigma^{ss}(k;\textbf{r})= \overline{n}_b
S(k;\beta,\overline{n}_b)$, with $S(k;\beta,\overline{n}_b)$ being
the equilibrium static structure factor of the homogeneous system.
Under these conditions (i.e., full equilibration and spatial
uniformity), from Eqs. (\ref{fluct5p})-(\ref{dzdtp}) we recover the
equilibrium version of the SCGLE theory \cite{scgle1,todos2}, and
from Eqs. (\ref{fdkinf})-(\ref{nep5pp}) we recover the corresponding
so-called bifurcation equations \cite{todos2}, using the terminology
of MCT \cite{goetze1}.

The full non-equilibrium SCGLE equations, Eqs.
(\ref{difeqdl})-(\ref{nep5pp}), can be solved only after several
elements have been specified. The most basic of them refers to the
nature of the system, defined by the pair interaction potential
$u(r)$, which determines the non-ideal contribution to the
electrochemical potential. This contribution is represented by the
term $-c[{\bf r};n]$ of the chemical equation of state, written as
$\beta\mu [{\bf r};n] = \beta\mu^{*}(\beta) + \ln n({\bf r}) -c[{\bf
r};n] + \beta \psi({\bf r})$. The functional dependence of $c[{\bf
r};n]$ on the concentration profile $n({\bf r})$ is a second
fundamental element that must be specified. One possibility is to
propose a theoretical approximation for this dependence, which in
the language of density functional theory is actually equivalent to
proposing an approximate free energy functional. For example, within
the simplest approximation, referred to as the Debye-H\"uckel or
random phase approximation, $c[{\bf r};n]$ is written as
$c^{(RPA)}[{\bf r};n]=-\beta \int d^3\textbf{r}'
u(|\textbf{r}-\textbf{r}'|)n({\bf r}')$ (which also defines an
approximation for the thermodynamic matrix, namely,
$\mathcal{E}^{(RPA)}[{\bf r},{\bf r}';n]=
\delta(\textbf{r}-\textbf{r}')/n({\bf
r})+u(|\textbf{r}-\textbf{r}'|)$).

A third element to specify refers to the external fields and the
thermodynamic constraints to which the system is subjected. We have
assumed so far that the external fields are static and represented
by $\psi({\bf r})$, whereas the thermodynamic constraints consists
of keeping the temperature field uniform, $T({\bf r},t)= T (t)\
(=1/k_B\beta(t))$, but not necessarily constant. There is, however,
no fundamental reason why we have to restrict ourselves to these
conditions. In fact, the general equations of the NE-SCGLE theory
above can be used, within the range of validity of the underlying
assumptions, to describe the response of the system to prescribed
time-dependent external fields $\psi({\bf r},t)$ or programmed
thermal constraints described by the time-dependent temperature
$T(t)$. This would be done by just including this possible
time-dependence in Eq. (\ref{difeqdl}) through the electrochemical
potential $\mu [{\bf r},t;n] = \mu^{*}(T(t)) + k_BT(t)\ln n({\bf r})
-k_BT(t)c[{\bf r};n] +  \psi({\bf r},t)$. Most commonly, however, we
assume that such time-dependent fields and constraints could be used
to drive the system to a prescribed initial state, described by the
mean value $\overline{n}^{0}(\textbf{r})$ and covariance
$\sigma^{0}(k;\textbf{r})$, for then programming the field and the
temperature to remain constant afterward,  $\psi({\bf r},t)=
\psi({\bf r})$ and $T(t)=T$ for $t>0$. The present theory then
describes how the system relaxes to its final equilibrium state
whose mean profile and covariance are
$\overline{n}^{eq}(\textbf{r})$ and $\sigma^{eq}(k;\textbf{r})$.

Describing this response at the level of the mean local
concentration profile $\overline{n}(\textbf{r},t)$ is precisely the
aim of the recently-developed \emph{dynamic} density functional
theory (DDFT) \cite{tarazona1, archer}. To establish direct contact
with this theory, let us consider the limit in which we neglect the
friction effects embodied in $\Delta \zeta^*(\tau; \textbf{r},t)$ by
setting $b(\textbf{r},t)=1$ in our main equations, namely, Eqs.
(\ref{difeqdlp}) and (\ref{relsigmadif2p}). We notice that under
these conditions Eq. (\ref{difeqdlp}) corresponds to the central
equation of DDFT, which has been applied to a variety of systems,
including the description of the irreversible sedimentation of real
and simulated colloidal suspensions \cite{royalvanblaaderen}. We
should also mention that Tokuyama \cite{tokuyama1, tokuyama2} has
proposed an equation for the irreversible relaxation of
$\overline{n}(\textbf{r},t)$ which differs from such simplified
version of our Eq. (\ref{difeqdlp}) only in that it neglects
external forces as well as the effects of the interparticle direct
interactions embodied in the non-ideal part of the electrochemical
potential, i.e., it sets $c[{\bf r};n(t)]=0$ in Eq. (\ref{1}). In
contrast, Tokuyama's theory does include some effects of the direct
interparticle interactions, as well as of hydrodynamic interactions,
on the matrix $\mathcal{L} [\textbf{r},\textbf{r}';t]$ (see Eq.
(\ref{matrixl3})), through the replacement of the diffusion
coefficient $D^0$ by the short-time self diffusion coefficient
$D_S(\overline{n}({\bf r};t))$ that depends as an ordinary function
on the local concentration. Just like DDFT, Tokuyama's theory
provides a description of the spatially inhomogeneous relaxation of
the local concentration profile. Furthermore, it seems to predict
dynamic arrest for hard-sphere dispersions. The current versions of
dynamic density functional theory, on the other hand, cannot predict
dynamic arrest phenomena because of the simplifying approximation
$b(\textbf{r},t)=1$.

The theory proposed in the present work shares some elements with
both of these theoretical developments, in the sense that it is also
aimed at describing the non-equilibrium relaxation of the local
equilibrium profile. We consider, however, that the description of
the irreversible relaxation of the macroscopic state of the system
is not complete without the description of the relaxation of the
covariance matrix $\sigma(\textbf{r},\textbf{r}';t)$ in Eq.
(\ref{relsigmadif2}) (or (\ref{relsigmadif2p})) and without the
inclusion of the effects embodied in the local mobility function
$b(\textbf{r},t) \ne 1$, i.e., in the friction function $\Delta
\zeta^*(\tau; \textbf{r},t) \ne 0$. In this regard it is also
important to point out that in the limit $b(t) \to 1$ of  Eq.
(\ref{relsigmadif2p}) one can recognize an equation that has been
fundamental in the description of the early stage of spinodal
decomposition \cite{cook,langer,dhont,goryachev}. For example, with
the additional small-wave-vector approximation for
$\mathcal{E}^{eq}_f(k)$, namely, $\mathcal{E}^{eq}_f(k)\approx
\mathcal{E}_0 +\mathcal{E}_2 k^2 +\mathcal{E}_4k^4$, this equation
is employed in the description of the early stages of spinodal
decomposition (see, for example, Eq. (2.11) of Ref. \cite{langer},
in which $\mathcal{E}_4=0$, or Eq. (23) of Ref. \cite{goryachev}).

Another particular limiting condition that merits discussion, now in
the context of the complete theory, corresponds to the quasistatic
process, characterized by a trajectory
$\overline{n}^{l.e.}(\textbf{r},t)$ and
$\sigma^{l.e.}(k;\textbf{r},t)$ that satisfies what we refer to as
the local equilibrium approximation. In this idealized process the
system is driven from a given initial equilibrium state described by
$\overline{n}^{0}(\textbf{r})$ and $\sigma^{0}(k;\textbf{r})$ to a
final equilibrium state described by $\overline{n}^{eq}(\textbf{r})$
and $\sigma^{eq}(k;\textbf{r})$ by extremely slow time-dependent
fields and constraints in such a manner that $(\partial
\overline{n}^{l.e.}(\textbf{r},t)/\partial t)$ and $(\partial
\sigma^{l.e.}(k;\textbf{r},t)/\partial t)$ virtually vanish due to
the fact that at each time $t$ the system is allowed to
approximately attain the two equilibrium conditions, $\nabla
\mu[\textbf{r};\bar{n}^{l.e.}]=0$ and
$\sigma^{l.e.}(k;\textbf{r},t)=\mathcal{E}^{-1}(k;\overline{n}^{l.e.}(\textbf{r},t))$.
A quasistatic process, however, is an idealized and rather
unrealistic concept, at least in the limit of small wave-vectors, in
which the relaxation times diverge as $k^{-2}$ (see Eq.
(\ref{relsigmadif2p})). In fact, far more interesting is the
opposite limit, in which the system, initially at equilibrium with a
static field $\psi^{(0)}(\textbf{r})$ and temperature $T^{(0)}$,
must adjust itself in response to an instantaneous change of these
control parameters to new values $\psi^{(f)}(\textbf{r})$ and
$T^{(f)}$, according to the ``program" described by
$\psi(\textbf{r},t) = \psi^{(0)}(\textbf{r}) \theta
(-t)+\psi^{(f)}(\textbf{r})\theta (t)$ and $T(t) = T^{(0)}\theta
(-t)+T^{(f)}\theta (t)$ with $\theta (t)$ being Heavyside's step
function.

Under the conditions described by this instantaneous quench program
the predicted non-equilibrium trajectory
$\overline{n}(\textbf{r},t)$ and $\sigma(k;\textbf{r},t)$ will
spontaneously reach the new thermodynamic equilibrium state
$\overline{n}^{eq}(\textbf{r})$ and $\sigma^{eq}(k;\textbf{r})$,
unless dynamic arrest conditions arise along this non-equlibrium
trajectory. This is, of course, the most fascinating possibility,
and it was the main motivation to carry out the present
non-equilibrium extension of the SCGLE theory. A simple manner to
monitor if this possibility will actually interfere with the process
of full equilibration is to solve Eq. (\ref{nep5pp}) for the squared
localization length $\gamma^{eq}(\textbf{r})$ when
$\overline{n}(\textbf{r},t)$ and $\sigma(k;\textbf{r},t)$ are given
their expected equilibrium values $\overline{n}^{eq}(\textbf{r})$
and $\sigma^{eq}(k;\textbf{r})$. If the resulting value of the
dynamic order parameter $\gamma^{eq}(\textbf{r})$ turns out to be
finite for \textbf{r} in some portion of the system, we should
expected the system to become dynamically arrested at least in that
region. The possible scenarios in which this might be predicted to
occur are hidden in the full NE-SCGLE equations above and in the
specific systems and conditions that might be considered. In order
to explore how reasonable these expectations may be, in a separate
work \cite{aging2} we apply for the first time the full NE-SCGLE
theory above, to the quantitative description of the response of a
simple model glass-forming colloidal liquid subjected to a spatially
homogeneous instantaneous quench to conditions where dynamic arrest
is expected on the basis of the procedure just outlined.

\section{Discussion and summary}\label{VI}

In this paper we have proposed the extension of the self-consistent
generalized Langevin equation (SCGLE) theory of colloid dynamics to
general non-equilibrium conditions. This extension describes in
principle the process in which the spontaneous evolution of the
system towards its thermodynamic equilibrium state could be
interrupted by the appearance of conditions for dynamic arrest. The
main fundamental basis of this general self-consistent theory were
provided by the general principles of Onsager's theory of
equilibrium thermal fluctuations, or, better, by the extension of
Onsager's theory to non-stationary and non-Markovian conditions
\cite{generalizedonsager}, whose review was the subject of section
\ref{II}. Clearly, this extended theory of irreversible processes is
in principle applicable to other relaxation phenomena outside the
realm of colloid dynamics.

The application of this extension of Onsager's theory to the
description of the irreversible evolution of the structure and
dynamics of a colloidal liquid was carried out in Sect. \ref{III}.
The resulting non-equilibrium theory of colloid dynamics contains as
particular cases a number of relevant limiting conditions. For
example, the evolution equation for the mean profile
$\overline{n}(\textbf{r},t)$ is found to contain the fundamental
equation of dynamic density functional theory as a particular limit,
whereas the basic equation employed to describe the evolution of the
static structure factor in the early stages of the process of
spinodal decomposition can be recognized as a particular limit of
the evolution equation for the covariance. The general theory,
however, also allows its application to the description of the
irreversible processes, such as aging, associated with dynamic
arrest transitions. In particular, it should in principle be
suitable to describe processes of dynamically arrested spinodal
decomposition.

Let us finally notice that the general non-equilibrium theory of
dynamic arrest has built in a very natural manner the description of
static and dynamic heterogeneities, since at any evolution time all
the relevant static and dynamic properties are defined at each point
in space, and cannot a priori be assumed spatially homogeneous. As a
zeroth order approximation, however, one may simplify the full
self-consistent theory assuming spatial homogeneity, as it is done
in the accompanying paper \cite{aging2}, and for some purposes this
simplifying approximation may suffice to provide an acceptable
first-order scenario of important non-equilibrium processes.

Since the final value of this general theoretical proposal depends
on its actual predictive power, in the accompanying paper we
illustrate the practical and concrete use of the present
non-equilibrium theory with a quantitative application to the
prediction of the aging processes occurring in a suddenly quenched
colloidal liquid, whose static structure factor and its van Hove
function evolve irreversibly from the initial conditions before the
quench to a final, dynamically arrested state. As reported there,
the comparison of the corresponding numerical results with available
simulation data seem highly encouraging.

\bigskip

ACKNOWLEDGMENTS: We dedicate this paper to the memory of Joel
Keizer, whose ideas provided a continuous and invaluable guidance.
The authors also acknowledge Rigoberto Ju\'arez-Maldonado, Alejandro
Vizcarra-Rend\'on and Luis Enrique S\'anchez-D\'iaz for stimulating
discussions and for their continued interest in this subject. This
work was supported by the Consejo Nacional de Ciencia y
Tecnolog\'{\i}a (CONACYT, M\'{e}xico), through grants No. 84076 and
CB-2006-C01-60064, and by Fondo Mixto CONACyT-SLP through grant
FMSLP-2008-C02-107543.

\vskip1cm

\appendix{\bf {Appendix A. FUNDAMENTAL THERMODYNAMIC
FRAMEWORK}}\label{A}

This appendix summarizes the essential concepts of the thermodynamic
theory of inhomogeneous fluids as a straightforward application of
the first and second laws of classical thermodynamics \cite{callen}
to a system that cannot be spatially homogeneous. Augmented with
elementary concepts of the thermodynamic theory of fluctuations
\cite{keizer,landaulifshitz,callen,greenecallen}, the resulting
purely phenomenological description involves some basic equations
that also appear in microscopic statistical mechanical theories,
such as density functional theory of inhomogeneous fluids
\cite{evans}.

The first law of thermodynamics states that, in a macroscopic system
formed by $N$ particles in a volume $V$, the total internal energy
$E$ is a state function, whereas the second law postulates the
existence of the entropy $S$,  another state function with the
property that a closed system will spontaneously search for the
state with the maximum $S$, and this state is referred to as the
thermodynamic equilibrium state. The functional relationship between
the entropy and the other $\emph{extensive}$ variables $E$, $N$, and
$V$ is referred to as the \emph{fundamental thermodynamic relation
(FTR)} of the system, written as $S=S(E,N,V)$. The presence of
external fields, however, may cause spatial inhomogeneities in the
distribution of matter and energy. The description of the possible
thermodynamic states of this system then requires of more
information than that contained in the value of the total properties
$E$, $N$, and $V$.

\subsection {Thermodynamic state space of a non-uniform
system.}

For this reason we mentally partition the volume $V$ in a number $C$
of smaller portions (or {\it cells}), whose internal energy,
particle number, and volume, we denote by $E^{(r)}, N^{(r)}$ and
$V^{(r)}$, respectively, with $r=1,2,...,C.$ Then, the fundamental
thermodynamic relation of this system reads $S=S[{\bf E,N,V}]$,
where {\bf E,\ N} and {\bf V} are \emph{C}-dimensional vectors with
components $E^{(r)},N^{(r)}$, and $V^{(r)}$ $(r=1,2,...,C).$ For the
sake of simplicity let us assume that the volumes $V^{(r)}$ are all
equal, $V^{(r)}=\Delta V=V/C$, and remain fixed, so that only the
variables [{\bf E,N}] are needed to define a thermodynamic state.
Specific values given to each of the components $N^{(r)}$ of the
vector {\bf N} define a \emph{particle number profile}, and specific
values of the components $E^{(r)}$ define a specific \emph{energy
profile }{\bf E}. For notational convenience let us also introduce
the $M$-component vector {\bf a} (with $M=2C$) as ${\bf a}\equiv
[{\bf E}, {\bf N}]$. Then, a specific particle number profile {\bf
N} and a specific energy profile {\bf E} define a specific
\emph{thermodynamic profile} {\bf a}. The set of all possible
thermodynamic profiles, that we refer to as the \emph{entire
thermodynamic state space} $\mathcal{T}$, is then identical to the
set of all particle and energy profiles that result from giving the
components $E^{(r)}, N^{(r)}$ any value in the range $0\le E^{(r)}<
\infty$ and $0\le N^{(r)} < \infty$. The fundamental thermodynamic
relation of this system, which then reads

\begin{equation}
S=S[{\bf E,N}], \label{1.4}
\end{equation}
assigns a value of the entropy to any possible thermodynamic profile
$[{\bf E,N}]={\bf a}\in \mathcal{T}$.

None of the elements of the state space $\mathcal{T}$ of a given
system is a priori an equilibrium or non-equilibrium state. The
second law of thermodynamics can distinguish which element of
$\mathcal{T}$ is the thermodynamic equilibrium state only after
specifying \emph{i}) the system, \emph{ii}) the \emph{external
fields} acting on its constituent particles and \emph{iii}) the
global thermodynamic constraints (such as isolation or contact with
reservoirs) imposed on the system. The system is defined by
specifying the pair interaction energy $u^{(r,r')}$ between two of
its particles located at cells $r$ and $r'$. The given array of
external fields acting on the particles is described by the
corresponding total potential energy of one particle at cell $r$ ,
that we shall denote by $\psi^{(r)}$. We shall refer to the
\emph{C}-dimensional vector $\Psi$ with components $\psi^{(r)}$
($r=1,2,...,C,$) as an \emph{external potential profile}.

The conceptually simplest and most important \emph{global}
thermodynamic constraint that may be imposed on the system is total
isolation, which prevents the system (of fixed total volume $V$)
from exchanging matter and energy with external reservoirs. Thus,
the total energy $E$ and particle number $N$ are constant,

\begin{equation}
\sum_{r=1}^CE^{(r)}  =
E\ \ \ \ (=const.)  \label{1.5a}
\end{equation}
and
\begin{equation}
 \sum_{r=1}^CN^{(r)}  =  N\ \ \ \ (=const.).
\label{1.5b}
\end{equation}
We may classify the elements of the entire thermodynamic state space
$\mathcal{T}$ of a given system according to the possible closure
conditions, i.e., according to the specific values of $(E,N,V)$.
Thus, each specific value of $(E,N,V)$ defines a specific subspace
$\tau(E,N,V) \subset \mathcal{T}$, which contains all the
thermodynamic profiles ${\bf a}=[{\bf E,N}]$ consistent with the
referred isolation condition. We may then say that any two subspaces
$\tau(E,N,V)$ and $\tau(E',N',V')$ are disjoint unless $E=E'$,
$N=N'$, and $V=V'$, and that the union of the subspaces
$\tau(E,N,V)$ for all possible values of $(E,N,V)$ is identical to
the entire thermodynamic state space $\mathcal{T}$. The second law
of thermodynamics then states that, for a fixed external potential
profile $\Psi$, an isolated system will spontaneously relax from any
arbitrary thermodynamic profile $[{\bf E,N}] \in \tau(E,N,V)$
towards the particular profile [${\bf E}_{eq}, {\bf N}_{eq}$] that
maximizes the entropy within the subspace $\tau(E,N,V)$. This means
that each possible profile $\Psi$ will identify a member of
$\tau(E,N,V)$ as ``its" corresponding equilibrium profile [${\bf
E}_{eq}, {\bf N}_{eq}$].

\subsection{Equations of state and conditions for thermodynamic equilibrium.}

The fundamental thermodynamic relation $S=S[{\bf E,N}]$ can also be
written in its differential form as
\begin{equation}
dS[{\bf E,N}]/k_B=\sum_{r=1}^C\beta^{(r)}[{\bf E,N}]dE^{(r)}-
\sum_{r=1}^C\beta\mu^{(r)}[{\bf E,N}]dN^{(r)}, \label{1.6}
\end{equation}
where  ``$\beta^{(r)}$" and ``$\beta\mu^{(r)}$" denote the functions
of the variables {\bf E} and {\bf N} defined as
\begin{equation}
\beta^{(r)}= \beta^{(r)}[{\bf E,N}]\equiv
{\partial(S[{\bf E,N}]/k_B) \over \partial E^{(r)}}
\label{1.8a}\end{equation}

\begin{equation}
\beta \mu^{(r)}=\beta \mu^{(r)}[{\bf E,N}] \equiv
-{\partial[S({\bf E,N}]/k_B) \over \partial
N^{(r)}}. \label{1.8b}
\end{equation}
For simplicity we may denote by $\widetilde{\beta}$ the
$C$-dimensional vector with components $\beta^{(r)}$, and by
$\widetilde{\beta\mu}$ the C-dimensional vector with components
$\beta\mu^{(r)}$ so that, for example, Eq. (\ref{1.6}) can also be
written as $dS[{\bf E,N}]/k_B=\widetilde{\beta}[{\bf E,N}]\cdot
d\textbf{E}- \widetilde{\beta\mu}[{\bf E,N}]\cdot d\textbf{N}$. Eqs.
(\ref{1.8a}) and (\ref{1.8b}) are, respectively, the \emph{thermal }
and the \emph{chemical} equations of state \cite{callen}.
Equilibrium states will satisfy the extremum condition $dS[{\bf
E}_{eq}, {\bf N}_{eq}]=0$ which, together with  Eqs. (\ref{1.5a})
and (\ref{1.5b}), leads to the following set of $2C$ equations for
the $2C$ variables $[{\bf E}_{eq}, {\bf N}_{eq}]$
\begin{equation}
 \beta^{(r)}[{\bf E}_{eq}, {\bf N}_{eq}] =
\beta\ \ \ (=const.)
\label{1.7a}
\end{equation}

\begin{equation}
 \beta\mu^{(r)}[{\bf
E}_{eq}, {\bf N}_{eq}] =
\beta\mu \ \ \ (=const)
\label{1.7b}
\end{equation}
for $r=1,2,...,C$. Clearly, these are
merely the conditions for {\it internal} thermodynamic equilibrium,
which require that the intensive parameters do not vary from cell to
cell.

\subsection{Thermodynamic theory of fluctuations: covariance and
stability matrices.}

The equilibrium value of the thermodynamic profile
$\textbf{a}_{eq}\equiv [{\bf E}_{eq}, {\bf N}_{eq}]$ is then the
solution of the extremum condition in Eqs. (\ref{1.7a}) and
(\ref{1.7b}). There are, however, instantaneous departures from such
an equilibrium profile, whose properties can only be described in
statistical terms. This then means that the thermodynamic profile
$\textbf{a}=[{\bf E}, {\bf N}]$ \emph{must }be regarded as a
$M$-component random vector, subject to a probability distribution
$P^{eq}[{\bf a}]$ whose mean value $\overline{\textbf{a}}$ is the
equilibrium value $\textbf{a}^{eq}$. Thus, we must now recognize
that the macroscopic state of our system cannot be described simply
by indicating the mean value
$\overline{\textbf{a}}=\textbf{a}_{eq}$; instead, it must be
described by the full probability distribution function $P[{\bf a}]$
given, according to the thermodynamic theory theory of fluctuations
\cite{keizer,landaulifshitz,callen,greenecallen}, by the {\it
Boltzmann-Planck} expression $P^{eq}[{\bf
a}]=\exp\left[{\left(S[{\bf a}]-S[{\bf a}^{eq}]\right)/k_B}\right]$.
The $M$x$M$ covariance matrix
$\sigma^{eq}_{ij}\equiv\overline{(\delta {\bf a})( \delta {\bf
a})^\dagger}$ of this distribution function, with elements defined
as

\begin{equation}  \sigma^{eq}_{ij}= \overline{\delta a_i\delta a_j} \equiv \sum_{\bf a} P^{eq}[{\bf a}] (a_i-
a_i^{eq})(a_j-a_j^{eq}), \ \ \ \ \ i,j=1,2,...,M, \label{3.5}
\end{equation}
is given by the following exact and general result,

\begin{equation}
 \sigma^{eq}\cdot {\bf \mathcal E}^{eq}={\bf I} \label{3.7}
\end{equation}

\noindent where {\bf I} is the $M\times M$ identity matrix and ${\bf
\mathcal E}^{eq}$ is the equilibrium stability matrix, defined as
\begin{equation}
{\mathcal E}^{eq}_{ij} \equiv
-{1 \over k_B}\left({ \partial^2 S[{\bf a}] \over \partial
a_i \partial a_j}\right)_{{\bf a}={\bf a}^{eq}} . \label{3.6}
\end{equation}

At any arbitrary state ${\bf a}$ (not necessarily an equilibrium
state) one can define the second differential of theentropy as
$d^2S[{\bf a}]/k_B = -d{\bf a}^\dagger\cdot {\bf \mathcal E}[{\bf
a}]\cdot d{\bf a}$, with  ${\bf \mathcal E}[{\bf a}]$ being the
$M\times M$ matrix defined as
\begin{equation}
{\bf \mathcal E}_{ij}[{\bf a}] \equiv
-{1 \over k_B}\left({ \partial^2 S[{\bf a}] \over \partial
a_i \partial a_j}\right)    = -{1 \over k_B}\left({ \partial
F_i[{\bf a}] \over \partial a_j}\right). \label{3.9}
\end{equation}
In the last member of this equation, $F_{i}[{\bf a}]$ is the
thermodynamically conjugate variable of the extensive variable
$a_{i}$, defined as $F_{i}[{\bf a}]= \left(\partial S[{\bf
a}]/\partial a_i \right)$. The conjugate variables $F_{i}[{\bf a}]$
and the thermodynamic matrix ${\bf \mathcal E}_{ij}[{\bf a}]$, are
thus defined at any thermodynamic state {\bf a}. It is, however,
only when the state {\bf a} is an equilibrium state that these state
functions have an important extremum and stability significance. In
particular, it is only under conditions of thermodynamic equilibrium
that the matrix ${\bf \mathcal E}$ is the inverse of the covariance
matrix, according to Eqs. (\ref{3.7}) and (\ref{3.6}).

\subsection{Legendre-transformed fundamental thermodynamic
relation.}

Just like in ordinary classical thermodynamics, under some
circumstances one may prefer to express all the previous results not
in terms of $[{\bf E}, {\bf N}]$ as independent state variables, but
in terms of $[\widetilde{\beta}, {\bf N}]$. Regarding the internal
equilibrium conditions, Eqs. (\ref{1.7a})-(\ref{1.7b}), this amounts
to eliminate the variables $E^{(r)}$ from this set of 2C equations
by first solving the thermal-equilibrium condition $\beta^{(r)}[{\bf
E,N}]=\beta^R \label{2.2}$ for {\bf E}, and then substituting the
solution, denoted as $E_{eq}^{(r)}[\beta^R,{\bf N}]$, in Eq.
(\ref{1.7b}). This leads to C equations for $N^{(r)}\
(r=1,2,...,c)$, namely, $\beta\mu^{(r)}[\beta^R;{\bf
N}]\equiv\beta\mu^{(r)}[{\bf E}_{eq}[\beta^R;{\bf N}],{\bf
N}]=\beta^R\mu^R \ \ \ (r=1,2,...,C)$, where the functions
$\beta\mu^{(r)}[\beta;{\bf N}]$ and $\beta\mu^{(r)}[{\bf E, N}]$
differ from each other in the set of variables they depend on. This
procedure is done more formally by defining the Legendre
transformation of the fundamental thermodynamic relation $S=S[{\bf
E}, {\bf N}]$, which reads
$\mathcal{F}=\mathcal{F}[\widetilde{\beta}, {\bf N}]\equiv S[{\bf
E}, {\bf N}]/k_B -\widetilde{\beta}\cdot \textbf{E}$ . The new
``thermodynamic potential" $\mathcal{F}[\widetilde{\beta}, {\bf N}]$
now plays the role of the entropy $S$ but in the thermodynamic state
space spanned by the variables $[\widetilde{\beta}, {\bf N}]$, and
is related with the Helmholtz free energy $\mathcal{A}$ by
$\mathcal{F}[\widetilde{\beta}, {\bf N}]=-\beta\mathcal{A}$.

This representation is most convenient under conditions in which the
$N$ particle system is in contact with a thermal reservoir (in our
case the supporting solvent) that keeps temperature constant and
uniform, in which the case system is constrained to the
thermodynamic state subspace $\mathcal{T}(\beta^R;N,V)$ defined by
$[\widetilde{\beta}_{eq}, {\bf N}]$, with $\beta^{(r)}_{eq}=\beta^R$
for all cells $r$.  This means that within this constrained
thermodynamic subspace the FTR can be written in its differential
form as $d\mathcal{F}=-\beta^R\widetilde{\mu}[\beta^R;{\bf N}]\cdot
d\textbf{N}$ and the chemical equation of state
$\mu^{(r)}=\mu^{(r)}[\beta^R;{\bf N}]$ now expresses the components
of the vector $\widetilde{\mu}$ as a function of the profile
\textbf{N} and of the thermal reservoir parameter $\beta^R$, which
may then be considered a control parameter. The main advantage of
this representation is the simplification of the equilibrium
conditions in Eqs. (\ref{1.7a}), (\ref{1.7b}). Thus, the equilibrium
concentration profile ${\bf N}_{eq}$ is now determined by the
condition that the electrochemical potential remains spatially
uniform, i.e.,
\begin{equation}
 \beta\mu^{(r)}[{\bf N}_{eq};\beta] =
\beta\mu \ \ \ (=const).
\label{chemequil}
\end{equation}

This representation also simplifies the discussion of the
fluctuations $\delta \textbf{N}$ around the equilibrium
concentration profile $\textbf{N}_{eq}$. To see this, let us write
Eq. (\ref{3.7}) more explicitly as

\begin{equation} \left[
  \begin{array}{ c c }
     \overline{\delta \textbf{E} \delta \textbf{E}^\dagger} &\overline{\delta \textbf{E} \delta \textbf{N}^\dagger} \\
                 \overline{\delta \textbf{N} \delta \textbf{E}^\dagger} &\overline{\delta \textbf{N} \delta
                 \textbf{N}^\dagger}
  \end{array} \right]
 \left[
  \begin{array}{ c c }-\left({\partial  \widetilde{\beta}[{\bf E}, {\bf N}] \over \partial \textbf{E}}\right) &
-\left({\partial  \widetilde{\beta}[{\bf E}, {\bf N}]  \over \partial \textbf{N}}\right) \\
                 \left( {\partial  \widetilde{\beta\mu} [{\bf E}, {\bf N}] \over  \partial \textbf{E}}\right) &
\left({\partial  \widetilde{\beta  \mu}[{\bf E}, {\bf N}] \over
\partial \textbf{N}}\right)
  \end{array} \right]_{(eq)}=
  \left[\begin{array}{c c}\textbf{I}&\textbf{0}\\ \textbf{0}&\textbf{I}\end{array}\right]\label{3.11pp}
\end{equation}

\noindent where the subindex $``(eq)"$ means that the thermodynamic
derivatives in this equation are evaluated at $[{\bf E}, {\bf N}] =
[{\bf E}_{eq}, {\bf N}_{eq}]$. By inverting the thermodynamic
matrix, along with some straightforward thermodynamic algebra, one
can show that the covariance $\overline{\delta \textbf{N} \delta
\textbf{N}^\dagger}$ satisfies
\begin{equation}   \overline{\delta \textbf{N} \delta \textbf{N}^\dagger} \cdot
\left({\partial \widetilde{\beta \mu} [\textbf{N};\beta] \over
\partial \textbf{N}}\right)_{\textbf{N}=\textbf{N}_{eq}}  = \textbf{I}.
\label{covariance}
\end{equation}
This result, however, is again Eq. (\ref{3.7}) with
$\textbf{a}=\textbf{N}$ and corresponding to the global constraint
of contact with a thermal reservoir that keeps temperature constant
and uniform.


\bigskip

\begin{thebibliography}{99}

\bibitem{scgle1}  L. Yeomans-Reyna and M. Medina-Noyola, Phys. Rev. E {\bf
64}, 066114 (2001).

\bibitem{marco2} M. A. Ch\'avez-Rojo and M. Medina-Noyola, Phys. Rev. E {\bf 72}, 031107
(2005); ibid {\bf 76}: 039902 (2007).

\bibitem{rmf}  P.E. Ram\'{\i}rez-Gonz\'alez {\it et al.}, Rev. Mex. F\'{\i}sica
\textbf{53}, 327  (2007).

\bibitem{todos2} R. Ju\'arez-Maldonado {\it et al.}, Phys. Rev. E {\bf 76}, 062502 (2007).

\bibitem{pham1} K. N. Pham, S. U. Egelhaaf, P. N. Pusey, and W. C. K. Poon,
Phys. Rev. E \textbf{69}, 011503 (2004).

\bibitem{cipelletti1} L. Cipelletti and L. Ramos, J. Phys.: Condens. Matter \textbf{17},
\textbf{253}, 285 (2005).

\bibitem{martinezvanmegen} V. A. Martinez, G. Bryant, and W. van Megen, Phys. Rev. Lett.
\textbf{101}, 135702 (2008).

\bibitem{sanz1} E. Sanz et al., J. Phys. Chem. B {\bf 112}, 10861 (2008).

\bibitem{lu1} P. J. Lu et al., Nature 453: 499 (2008).

\bibitem{tarazona1} U. Marini Bettolo Marconi and P. Tarazona, J. Chem. Phys. \textbf{110}, 8032
(1999); ibid., J. Phys.: Condens. Matter \textbf{12}, A413 (2000)

\bibitem{langer} J. S. Langer, M. Bar-on, and H. D. Miller, Phys. Rev. A \textbf{11}, 1417 (1975).

\bibitem{aging2} P. E. Ram\'{\i}rez-Gonz\'alez and M. Medina-Noyola,
Phys. Rev. E (2010, submitted).


\bibitem{1}  P. N. Pusey, in {\em Liquids, Freezing and the Glass transition},
edited by J. P. Hansen, D. Levesque, and J. Zinn-Justin (Elsevier,
Amsterdam, 1991).

\bibitem{6}  W. Hess and R. Klein, Adv. Phys. {\bf 32}, 173 (1983).

\bibitem{2}  G. N\"{a}gele, Phys. Rep. {\bf 272}, 215 (1996).

\bibitem{mcquarrie}  D. A. McQuarrie {\em Statistical Mechanics}, Harper \& Row (New York, 1973).

\bibitem{13}  N. J. Wagner {\em Phys. Rev. E}, {\bf 49}, 376 (1994).



\bibitem{scgle0}  L. Yeomans-Reyna and M. Medina-Noyola, Phys. Rev. E {\bf
62}, 3382 (2000).

\bibitem{scgle2}  L. Yeomans-Reyna, H. Acu\~{n}a-Campa,
F. Guevara-Rodr\'{\i}guez, and M. Medina-Noyola, Phys. Rev. E {\bf
67}, 021108 (2003).

\bibitem{marco1} M. A. Ch\'avez-Rojo and M. Medina-Noyola, Physica A {\bf 366}, 55 (2006).

\bibitem{todos1} L. Yeomans-Reyna {\it et al.},  Phys. Rev. E \textbf{76},
041504 (2007).

\bibitem{attractive1} P. E. Ram\'irez-Gonz\'alez  {\it et al.}, J. Phys.: Cond.
Matter, {\bf 20}: 20510 (2008).

\bibitem{soft1} P. E. Ram\'irez-Gonz\'alez  and M. Medina-Noyola, J. Phys.: Cond. Matter, {\bf 21}, 75101 (2009).

\bibitem{rigo1} R. Ju\'arez-Maldonado and M. Medina-Noyola, Phys. Rev. E {\bf 77}, 051503 (2008).

\bibitem{rigo2} R. Ju\'arez-Maldonado and M. Medina-Noyola, Phys. Rev. Lett. \textbf{101}, 267801 (2008).

\bibitem{luis1} L. E. S\'anchez-D\'iaz, A. Vizcarra-Rend\'on, and R. Ju\'arez-Maldonado, Phys. Rev.
Lett. \textbf{103}, 035701 (2009).


\bibitem{angell} C. A. Angell, Science {\bf 267}, 1924 (1995).

\bibitem{debenedetti} P. G. Debenedetti and F. H. Stillinger, Nature {\bf 410}, 359 (2001).

\bibitem{goetze1}  W. G\"{o}tze, in {\em Liquids, Freezing and Glass Transition},
edited by J. P. Hansen, D. Levesque, and J. Zinn-Justin
(North-Holland, Amsterdam, 1991).

\bibitem{goetze2} W. G\"{o}tze and L. Sj\"ogren, Rep. Prog. Phys. {\bf
55}, 241 (1992).

\bibitem{goetze3}  W. G\"{o}tze and E. Leutheusser, Phys. Rev. A {\bf 11}, 2173 (1975).

\bibitem{goetze4} W. G\"{o}tze, E. Leutheusser and S. Yip, Phys. Rev. A {\bf
23}, 2634 (1981).

\bibitem{vanmegen1} W. van Megen and P. N. Pusey, Phys. Rev. A {\bf
43}, 5429 (1991).

\bibitem{bartsch} E. Bartsch et al., J. Chem. Phys. {\bf 106}, 3743 (1997).

\bibitem{beck} C. Beck. W. H\"artl, and R. Hempelmann,  J. Chem. Phys. {\bf 111}, 8209 (1999).

\bibitem{chen1} S.-H. Chen {\it et al.}, Science {\bf 300}, 619 (2003).

\bibitem{chen2} W. R. Chen {\it et al.}, Phys. Rev. E {\bf 68}, 041402
(2003).

\bibitem{pham} K. N. Pham et al., Science, {\bf 296}, 104 (2002).

\bibitem{sciortinotartaglia} F. Sciortino and P. Tartaglia, Adv. Phys. {\bf 54}, 471 (2005).

\bibitem{buzzacaro} S. Buzzaccaro et al., Phys. Rev. Lett. \textbf{99}, 098301 (2007)





\bibitem{cugliandolo1} L. F. Cugliandolo and J. Kurchan, Phys. Rev. Lett. \textbf{71}, 173
(1993).

\bibitem{kobbarrat} W. Kob and J.-L. Barrat, Phys. Rev. Lett. \textbf{78}, 4581 (1997).

\bibitem{foffiaging1} G. Foffi, E. Zaccarelli, S. Buldyrev, F. Sciortino, and P. Tartaglia,
{\em J. Chem. Phys.} {\bf 120}, 8824 (2004).

\bibitem{puertas1} A. M. Puertas, M Fuchs, and M. E. Cates, Phys. Rev. E \textbf{75}, 031401
(2007).

\bibitem{latz} A. Latz, J. Phys.: Condens. Matter, \textbf{12} (2000) 6353.



\bibitem{berne} B. Berne, ``Projection Operator
Techniques in the theory of fluctuations", in \emph{Statistical
Mechanics, Part B: Time-dependent Processes}, B. Berne, ed. (Plenum,
New York, 1977).

\bibitem{degregorio} P. De Gregorio et al., Physica A, \textbf{307}, 15 (2002).

\bibitem{keizer} J. Keizer, \emph{Statistical Thermodynamics of Nonequilibrium Processes}, Springer-Verlag (1987).

\bibitem{degrootmazur} S. R. de Groot and P. Mazur \emph{Non-equlibrium Thermodynamics}, Dover, New York (1984).

\bibitem{casasvazquez0} G. Lebon, D. Jou, and J. Casas-V\'azquez, \emph{Understanding
Non-equilibrium Thermodynamics Foundations, Applications,
Frontiers}, Springer-Verlag Berlin Heidelberg (2008).



\bibitem{onsager1} L. Onsager, Phys. Rev. \textbf{37}, 405 (1931).

\bibitem{onsager2} L. Onsager, Phys. Rev. \textbf{38}, 2265 (1931).

\bibitem{onsagermachlup1} L. Onsager and S. Machlup, Phys. Rev. \textbf{91}, 1505 (1953).

\bibitem{onsagermachlup2} S. Machlup and L. Onsager, Phys. Rev. \textbf{91}, 1512 (1953).

\bibitem{delrio}  M. Medina-Noyola and J. L. del R\'\i o-Correa, {\em Physica
146A}, 483 (1987).

\bibitem{faraday}  M. Medina-Noyola, {\em Faraday Discuss. Chem. Soc.} {\bf 83},
21 (1987).

\bibitem{generalizedonsager} M. Medina-Noyola, arXiv:0908.0521v1 [cond-mat.stat-mech] 4 Aug
2009.

\bibitem{callen} H. Callen, \emph{Thermodynamics}, John Wiley, New York(1960).

\bibitem{evans} R. Evans, Adv. Phys. {\bf 28}: 143(1979).

\bibitem{archer} A. J. Archer and M. Rauscher, J. Phys. A \textbf{37}, 9325 (2004).

\bibitem{royalvanblaaderen} C. P. Royall et al.,
Phys. Rev. Lett. \textbf{98}, 188304 (2007).

\bibitem{landaulifshitz} L. D. Landau and E. M. Lifshitz,  \emph{Statistical Physics}, Addison-Wesley, Reading (1974).


\bibitem{greenecallen} R. F. Greene and H. B. Callen, Phys. Rev.
\textbf{83}: 1231 (1951)

\bibitem{tokuyama1} M. Tokuyama, Phys. Rev. E \textbf{54}, R1062 (1996).

\bibitem{tokuyama2} M. Tokuyama, Y. Enomoto, and I. Oppenheim, Phys. Rev. E \textbf{56}, 2302 (1997).




\bibitem{cook} H. E. Cook, Acta Metall. \textbf{18}, 297 (1970).

\bibitem{dhont} J. K. G. Dhont, J. Chem. Phys. \textbf{105}, 5112 (1996).

\bibitem{goryachev} S. B. Goryachev, Phys. Rev. Lett. \textbf{72}, 1850 (1994).


\end{thebibliography}
\end{document}